\documentclass[notitlepage,nofootinbib,superscriptaddress]{revtex4}

\usepackage{graphicx}
\usepackage{natbib}
\usepackage{latexsym}
\usepackage{graphics}
\usepackage{verbatim}
\usepackage{amssymb}
\usepackage{amsmath}
\usepackage{epsfig}
\usepackage{color}
\usepackage{rotating}
\usepackage{fancyhdr}
\usepackage{tabularx}
\usepackage{textcomp}  

\newcommand\Rey{\mbox{\textit{Re}}}  
\newcommand{\barf}[1]{\,\,\bar{\!\!{#1}}}

\newsavebox{\astrutbox}
\sbox{\astrutbox}{\rule[-5pt]{0pt}{20pt}}

\newcommand{\raux}{\mbox{$\tilde{r}$}}
\newcommand{\xaux}{\mbox{$\tilde{x}$}}

\newcommand{\ul}[1]{\underline{#1}}

\newcommand{\sst}[1]{\scriptscriptstyle{#1}}
\newcommand{\dsty}{\displaystyle}

\newcommand{\Nab}{\mbox{\boldmath$\nabla$}}
\newcommand{\Nabt}{\mbox{\boldmath$\tilde{\nabla}$}}

\newcommand{\vecdelta}{\mbox{\boldmath$\delta$}}
\newcommand{\veck}{\mbox{\boldmath$k$}}
\newcommand{\vecx}{\mbox{\boldmath$e$}_x}
\newcommand{\vecy}{\mbox{\boldmath$e$}_y}
\newcommand{\vecz}{\mbox{\boldmath$e$}_z}
\newcommand{\vecphi}{\mbox{\boldmath$e$}_{\phi}}
\newcommand{\vectheta}{\mbox{\boldmath$e$}_{\theta}}

\newcommand{\vecer}{\mbox{\boldmath$e$}_r}
\newcommand{\vecu}{\mbox{\boldmath$u$}}
\newcommand{\vecU}{\mbox{\boldmath$U$}}
\newcommand{\vecn}{\mbox{\boldmath$n$}}
\newcommand{\vecv}{\mbox{\boldmath$v$}}
\newcommand{\vecr}{\mbox{\boldmath$r$}}
\newcommand{\vecraux}{\mbox{\boldmath$\tilde{r}$}}
\newcommand{\vecf}{\mbox{\boldmath$f$}}
\newcommand{\vect}{\mbox{\boldmath$t$}}

\newcommand{\vecs}{\mbox{\boldmath$\sigma$}}

\newcommand{\vecF}{\mbox{\boldmath$F$}}
\newcommand{\vecA}{\mbox{\boldmath$A$}}
\newcommand{\vecB}{\mbox{\boldmath$B$}}
\newcommand{\vecC}{\mbox{\boldmath$C$}}

\newcommand{\vecK}{\mbox{\boldmath$K$}}
\newcommand{\vecR}{\mbox{\boldmath$R$}}
\newcommand{\vecV}{\mbox{\boldmath$V$}}
\newcommand{\vecG}{\mbox{\boldmath$\mathcal{G}$}}

\newcommand{\vecO}{\mbox{\boldmath$0$}}
\newcommand{\vecOm}{\mbox{\boldmath$\Omega$}}
\newcommand{\vecom}{\mbox{\boldmath$\omega$}}

\newcommand{\vecL}{\mbox{\boldmath$L$}}

\newcommand{\vecGamma}{\mbox{\boldmath$\Gamma$}}

\newcommand{\per}{\sst \perp}
\newcommand{\pa}{\sst \parallel}
\newcommand{\veps}{\varepsilon}
\newcommand{\eps}{\epsilon}
\newcommand{\icomp}{\mbox{i}}
\newcommand{\icomps}{\mbox{\scriptsize i}}
\newcommand{\expo}{\mbox{e}}

\begin{document}

\title{Asymmetric steady streaming as a mechanism for acoustic propulsion of rigid bodies} 

\author{Fran\c cois Nadal}
\affiliation{Commissariat \`a l'Energie Atomique, 33114 Le Barp, France}
\author{Eric Lauga}
\email{e.lauga@damtp.cam.ac.uk}
\affiliation{Department of Applied Mathematics and Theoretical Physics, University of Cambridge,
Center for Mathematical Sciences, Wilberforce Road, Cambridge CB3 OWA, United Kingdom}

\begin{abstract}

Recent experiments  showed that standing acoustic waves could be exploited  to  induce self-propulsion of rigid metallic particles in the 
direction perpendicular to the acoustic wave. We  propose in this paper a physical mechanism for these observations  based on the interplay 
between inertial forces in the fluid and the geometrical asymmetry of the particle shape. We consider an axisymmetric rigid  near-sphere 
oscillating in a quiescent fluid along a direction perpendicular  to its symmetry axis. The kinematics of oscillations can be either 
prescribed or can result dynamically from the presence of an external oscillating  velocity field. Steady streaming in the fluid, 
the inertial rectification of the time-periodic oscillating flow, generates steady stresses on the particle which, in general, do 
not average to zero, resulting in a finite propulsion speed along the axis of the symmetry of the particle and perpendicular to 
the oscillation direction.  Our derivation of the propulsion speed is obtained  at leading order in the Reynolds number and the 
deviation of the shape from that of a sphere. The results of our model are consistent with the experimental measurements, and more 
generally explains how time periodic forcing from an acoustic field can be harnessed to generate autonomous motion.

\end{abstract}

\maketitle


\section{Introduction}

The transport of synthetic micro- and nano-scale particles is a well-studied  field of research, starting with the first studies
on the effect of electric fields on colloidal suspensions in the 1920s. 
The topic has recently seen a  revival of activity, due in part to the possible biomedical
and environmental use of these devices \cite{Nelson2010}. Indeed,   small controlled bodies could be 
employed to achieve transport of cargo and drug delivery \cite{Sundararajan2008,Burdick2008}, analytical
sensing  in biological media \cite{Campuzano2011b,Wu2010}.  Furthermore, their fast  motion could also
be efficiently used to perform wastewater treatment \cite{Soler2013}. 

While deformable synthetic micro-swimmers \cite{Dreyfus2005} are of fundamental interest to mimic the locomotion of  real cellular  organisms    
\cite{Bray2000,Lighthill1975,Lighthill1976,Brennen1977,Lauga2009}, 
rigid synthetic micro- and nano-swimmers appear  to provide a more  practical alternative. A number of  different mechanisms have been proposed 
to achieve propulsion of small rigid objects, as recently  reviewed   by Ebbens \& Howse \cite{Ebbens2010}
and Wang et al.~\cite{Wang2013}. The propulsion  mechanisms can be sorted into two generic categories:  external  mechanisms, in which a 
directional field is used to drive the object, and  autonomous mechanisms, where the object performs a local conversion of the energy from an exterior 
source field. In the latter case, symmetry breaking of the particle itself (shape, composition) is usually  required to achieve propulsion.

External strategies typically lead  to a global motion of the assembly of micro particles.
For instance, applying an electric field on a suspension of charged spherical colloids in an electrolyte leads to a collective motion 
of the assembly parallel to the field lines,
a phenomenon known as electrophoresis \cite{Smoluchowsky1921}. Applying a non uniform electric field on dielectric uncharged spherical particles in an 
electyrolyte leads as well to an ensemble motion of the colloids parallel to the field lines (dielectrophoresis \cite{Pohl1978}). Rigid particles can 
also be propelled by the mean of magnetic fields. For example, a time-varying magnetic field can be used to actuate in rotation an helical (chiral) 
body \cite{Ghosh2009,Zhang2009,Zhang2010}. 

Whereas external control is convenient for targeting and navigation, autonomous strategies are more suitable for swarming and cleaning tasks.
In this case, particles show independent trajectories able to cover a given region of fluid in a limited amount
of time than unidirectional similar trajectories resulting from external driving.
Autonomous motion can be achieved by  methods which typically require a breaking of the symmetry of the particle (not a requirement in the case 
of external forcing). Catalytic bimetallic microrods can propel at high velocities ($\sim\,$10 $\mu$m$\,$s$^{-1}$) in a liquid medium by
self-generating local electric fields maintained by a local gradient of charged species (self-electrophoresis) 
\cite{Paxton2004,Ibele2007,Ebbens2011}. If the generated species is uncharged, the concentration gradient can also trigger a net motion 
of the particle through self-diffusiophoresis \cite{Pavlick2011,Pavlick2013,Cordova2008}. 
 Similarly, autonomous propulsion can  be achieved by taking advantage 
of self-thermophoresis effects \cite{Jiang2010,Baraban2012,Qian2013}. 
Self-electrophoresis and self-diffusiophoresis have the important drawbacks to be incompatible with biological media such as blood,
for these processes rely on the use of toxic fuels -- {\it e.g.} hydrogene peroxide \cite{Paxton2004,Ebbens2011},
hydrazine \cite{Ibele2007} in the case of self-electrophoresis or norborene in the case of self-diffusiophoresis \cite{Pavlick2011} -- 
and are inefficient in  high-ionic strength media. Self-thermophoresis requires temperature  differences of a few Kelvins which 
makes it difficult  to use for medical applications. 

As an alternative, acoustic fields are good candidates to enable autonomous propulsion in biocompatible media, as recently demonstrated experimentally
by Wang {\it et al.} \cite{Wang2012}. In that work, it was shown that micron size metallic 
and bimetallic rods located in the pressure nodal plane of a standing 
acoustic wave could undergo planar autonomous motion with speeds of up to  200 $\mu$m$\,$s$^{-1}$.
In this paper, we provide a  model for these experimental results.  Specifically, we propose  asymmetric steady fluid streaming as a generic physical  
mechanism   inducing the propulsion of rigid particles  in a standing acoustic wave.  This mechanism requires a shape asymmetry of the particle,  
does not involve any other physical process than pure Newtonian hydrodynamics (in particular, no chemical reaction), and takes its origin in the 
non-zero net forces induced in the fluid by inertia  under time-periodic forcing.

After drifting towards the pressure nodal plane  under the effect of the radiation
pressure \cite{Doinikov1994a,Mitri2009}, a rigid particle can be viewed as a  body oscillating 
in a uniform oscillating velocity field - note that this is does not hold in the general case
where the particle is located at an arbitrary $X$-position in the resonator (see section \ref{acoustic_field}).  If $K_{\sst 0}$ and $R_{\sst 0}$ refer respectively to the wavenumber of the acoustic radiation and the typical size of the particle, this  assumption of local uniformity of the field is justified provided that  $K_{\sst 0}R_{\sst 0} \ll 1$, a limit true in the experiments in Ref.~\cite{Wang2012}. The motion of the particle relative to the surrounding fluid leads then to an oscillating perturbative flow which can be computed in the framework of unsteady Stokes flows. Such a viscous flow, when coupled with itself through the convective term of the  Navier-Stokes equation, forces a steady flow (so-called steady streaming),
together with a flow at twice the original pulsation. If the particle has a non-spherical shape, the force coming from the integration
of the corresponding steady streaming stress over the surface of the particle will generically non cancel out, leading to propulsion. 
Critically, in the absence of inertia, no propulsion would be possible since the initial transverse oscillatory  motion is time-reversible.
The breaking of symmetry in the geometry is also indispensable and, as originally shown by Riley \cite{Riley1966}, the net force coming
from the integration of steady inertial stress (steady streaming stress) over the surface of an oscillating sphere is zero. 

In order to mathematically model this physical mechanism,  
we first consider the problem of an axisymmetric near-sphere oscillating in a prescribed fashion in the transverse direction
in a quiescent fluid. The particle is assumed to be force-free in the direction of its axis of symmetry. 
We start by a near sphere  of harmonic polar equation (i.e.~one whose shape differs from the sphere
by a cosine of small amplitude) before considering  an {arbitrary} axisymmetric shape.  The case of a free particle in an oscillating
uniform velocity field is then addressed as it corresponds to the experimental situation
in which the particle is trapped at the pressure node of a standing acoustic radiation.    The  problem is governed by two dimensionless
parameters: a shape parameter,quantifying the distance to a perfect sphere, and the Reynolds number. Our calculations will present  the derivation
of the propulsion speed at leading order in both, giving rise to a propulsive force on the order of {\it shape parameter$\times$Reynolds number}.
To perform the perturbation analysis, we  expand the fields in  Reynolds number
and to introduce the shape parameter at each separated order in Reynolds.

The paper is organized as follows. Section \ref{presentation} is devoted to the presentation of the problem.
Geometry, governing equations, and boundary conditions are detailed. Section \ref{inertial_effects}
is dedicated to the derivation of an integral expression of the first-order (in Reynolds) propulsion speed.
Zeroth and first-order (in Reynolds) problems are successively addressed. The full solution to the zeroth-order
-- transverse oscillation of a near-sphere in a purely viscous fluid -- is presented first. As we are interested
in the first-order (in Reynolds) propulsion speed rather than in the full first-order flow field, the latter is not
derived explicitly and instead, we use a suitable form of Lorenz's reciprocal theorem to establish an integral
expression of the propulsion speed  \cite{Ho1974}.   Results provided by the numerical
integration of the integral expression of the propulsion speed
are presented in section \ref{num_int}. We then use section \ref{acoustic_field} to address 
the dynamics of an axisymmetric near-sphere free to move in an uniform oscillating exterior velocity field.
We show in particular  that the zeroth-order (in Reynolds) rotational oscillation of the near-sphere
is of second-order (in shape perturbation number), so that the propulsion speed computed in the case of a non
rotating particle (section \ref{inertial_effects}) can be used as  is. We conclude the paper by a discuss of the numbers
predicted by the model in relation to the original experiment \cite{Wang2012}. In Appendix \ref{origin_position}, we
demonstrate that the calculated propulsion speed does not depend
on the choice of the origin of the coordinate system, a technical but important detail. As the integral form of the  propulsion speed involves the expression of the flow field induced by an oscillating sphere in a purely viscous fluid, we recall its expression  in Appendix \ref{oscillating_sphere}.  Some further technical details concerning the zeroth-order problem are given in Appendix \ref{transv_oscillation}. The use of the reciprocal theorem requires an auxiliary flow field. The characteristic of such a flow (axial translation of an axisymmetric solid body at constant speed in a purely viscous fluid) are given in Appendix \ref{axial_translation}. Finally, in Appendix \ref{lastApp} we discuss   the dipolar forces appearing when  the rigid particle is not located at a pressure node of the acoustic field.

\section{Problem formulation \label{presentation}}

\subsection{Geometry and kinematics \label{geo_kin}}

\begin{figure}
\scalebox{0.9}{\includegraphics{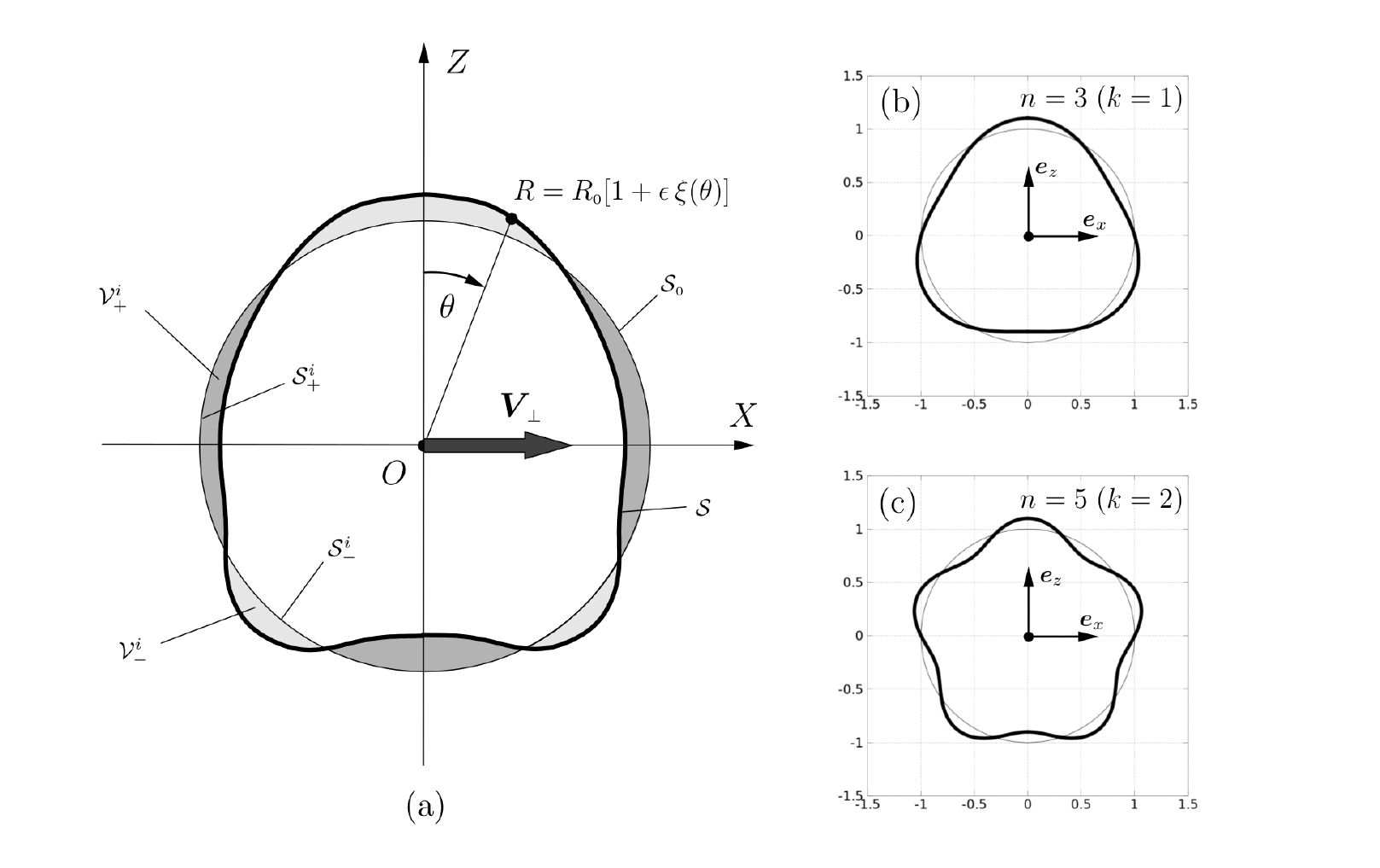}}
\caption{Geometry of the system (a). 
The shapes of solid bodies with symmetry $n=3$ and $n=5$
are displayed in figures (b) and (c) respectively.}
\label{geom}
\end{figure}

The setup of our calculation in shown in figure \ref{geom}. 
Both cartesian and spherical coordinate systems are used. Unit vectors of the cartesian (resp.~spherical) coordinate
system are referred to as $\vecx$, $\vecy$, and $\vecz$ (resp.~$\vecer$, $\vectheta$, and $\vecphi$). The position
is denoted by $\vecR$, and the spherical coordinates by $R$, $\theta$ and $\phi$.
We  use capital letters to refer to dimensional quantities,
force, position, and velocity variables. Corresponding dimensionless quantities are denoted
by small letters (this rule obviously does not apply for constants). 

We first consider an axisymmetric homogeneous solid body the axis of which is in the $z$-direction (in section \ref{acoustic_field}, 
the particle will be  free to rotate).  The body oscillates in a Newtonian fluid
(density $\rho$, viscosity $\mu$) along the transverse $x$-direction at frequency $\omega$. The amplitude
of its oscillations relative to the quiescent fluid is denoted $a$ such that the relative velocity of any point of the body is 
$\vecV_{\!\!\per} = \hat{\vecV}_{\!\!\per}\,\expo^{-\icomps\omega T}$, where $\hat{\vecV}_{\!\!\per} = a\,\omega\,\vecx$.

In order to allow an analytical solution, the solid body is assumed to take the form of a slightly deformed, axisymmetric sphere.
We thus write its shape as  
\begin{equation} 
R = R_{\sst 0}[1+\epsilon\,\xi(\theta)], \label{polar}
\end{equation}
where $R_{\sst 0}$ is the radius of the reference sphere,  $\eps \ll 1$ is the dimensionless small shape parameter and $\xi$ a
dimensionless function of order one. The surface of the
axisymmetric near-sphere is referred to as $\mathcal{S}$ and its volume is denoted by $\mathcal{V}_p$.
In our calculations we first assume that $\xi$ is of the form $\cos(n\theta)$, with $n = 2k+1$ ($k \ge 1$). The value $n=1$
is not considered since the corresponding body is 
equivalent to a sphere at order $O(\eps)$ (see Appendix \ref{origin_position}) and odd values of $n$ would lead to no
propulsion by symmetry. The case of an arbitrary (axisymmetric) shape is dealt with in section \ref{arbitrary_shape},
but we first perform the analysis for one of the terms of the Fourier expansion of the shape function susceptible to
provide a finite propulsion speed of the body along the direction of its axis of symmetry ($z$). Note that the function
$\xi(\theta) = \cos(n\theta)$ with $n = 2k+1$ satisfies the condition
\begin{equation}
\int_{{\mathcal{S}_{\sst 0}}}\!\!\xi\,d\mathcal{S} = 0, \label{cond_1}
\end{equation}
where $\mathcal{S}_{\sst 0}$ is the surface of sphere of radius $R_{\sst 0}$. Consequently, the sphere of radius
$R_{\sst 0}$ is the equivalent-volume sphere  and $\mathcal{V}_p = (4/3)\pi R_{\sst 0}^3$. 
Note also that the origin of the spherical coordinate system used in the paper is in general not the center of mass
of the body (except in section \ref{acoustic_field}),  and we have  thus that the equality
\begin{equation}
\int_{{\mathcal{S}_{\sst 0}}}\!\!\xi\, \vecn\,d\mathcal{S} = \vecO \label{cond_2}
\end{equation} 
is  not satisfied. This fact will be important when we address the translation/rotation coupled problem of the dynamics
of a near-sphere in a uniform exterior oscillating velocity field (section \ref{acoustic_field}).

\subsection{Governing equations and boundary conditions \label{governing_eq}}

The solid particle  is moving in the laboratory reference frame and  we choose to 
work in the frame of reference of the body. The dimensional velocity and pressure fields satisfy the incompressible Navier-Stokes equations
\begin{gather}
\frac{\partial\vecU}{\partial T} + (\vecU\cdot\Nab) \vecU = -\frac{1}{\rho}\Nab P + \nu \Nab^2 \vecU,\label{NS1_dim}\\
\Nab\cdot\vecU = 0 \label{NS2_dim},
\end{gather}
where $\nu = \mu/\rho$ is the kinematic viscosity of the fluid. In equation (\ref{NS1_dim}), the additional
inertial force field $- \rho \Gamma_{\sst O} = \icomp\,\rho \omega^2 a\,\expo^{-\icomps\,\omega t}\,\vecx$ due to
the acceleration $\Gamma_{\sst O}$ of the origin of the (non Galilean) reference frame 
has been incorporated in the pressure term (since this force field is the gradient
of the linear pressure field $\icomp\,\rho \omega^2 a\,\expo^{-\icomps\,\omega t}\,x$).
These equations can be made dimensionless by choosing $R_{\sst 0}$, $\hat{V}^{\per} = a\,\omega$,
$\omega^{-1}$, and $\mu \hat{V}^{\per}/R_{\sst 0}$ as typical length, velocity, time, and pressure scales, and one gets  
\begin{gather}
\lambda^2 \frac{\partial\vecu}{\partial t} + \Rey (\vecu\cdot\Nab) \vecu = -\Nab p + \Nab^2\vecu,\label{NS1_adim}\\
\Nab\cdot\vecu = 0 \label{NS2_adim}.
\end{gather}
In equation (\ref{NS1_adim}), $\lambda^{-1} = (\omega R_{\sst 0}^2/\nu)^{-1/2}$ quantifies the dimensionless distance over which
the vorticity diffuses, and $\Rey = R_{\sst 0}\hat{V}^{\per}/\nu = a R_{\sst 0} \omega/\nu$ is the Reynolds number. In the following,
$\lambda$ will be referred to as the viscous parameter. Due to the assumption $a\ll R_{\sst 0}$, the Reynolds
number is smaller than $\lambda^2$ by a factor $a/R_{\sst 0}$. Considering our choice of nondimensionalization, 
the polar equation of the surface is now written as
\begin{equation}
r = 1+\,\eps\,\xi(\theta). \label{polar_adim}
\end{equation}

Equations (\ref{NS1_adim}) and (\ref{NS2_adim}) have to be supplemented by a suitable set of boundary conditions.
We assume that, due to inertial effects, the force-free body will propel in the
$z$-direction (the only one allowed by symmetry) and the corresponding dimensionless propulsion
speed  is denoted $\vecv^{\pa}$.
As the analysis is performed in the reference frame of the particle, the boundary condition for the
velocity field then takes the following form
\begin{align}
&\vecu = \vecO\;\;\mbox{on $\mathcal{S}$}, \label{BC1}\\
&\vecu \rightarrow - \vecv^{\pa} - \vecv^{\per} = - v^{\pa}\,\vecz - \expo^{-\icomps t}\,\vecx,\;\;
\mbox{for}\;\;|\vecr| \rightarrow \infty. \label{BC2} 
\end{align}
With the aim of applying the reciprocal theorem (\S \ref{prop_speed}), 
we write the difference $\vecu' = \vecu + \vecv^{\pa}$, transforming equations (\ref{NS1_adim}) and (\ref{NS2_adim}) 
into new equations  as
\begin{gather}
\lambda^2 \frac{\partial\vecu'}{\partial t} + \Rey [(\vecu'\cdot\Nab) \vecu' - (\vecv^{\pa}\cdot\Nab) \vecu']
 = -\Nab p' + \Nab^2\vecu',\label{NS1_adim_2}\\
\Nab\cdot\vecu' = 0 \label{NS2_adim_2},
\end{gather}
where $p'=p$ since no additional pressure (stress) is associated with the uniform field $\vecv^{\pa}$.
The new set of boundary condition is
\begin{align}
&\vecu' = \vecv^{\pa}\;\;\mbox{on $\mathcal{S}$}, \label{BC1bis}\\
&\vecu' \rightarrow - \vecv^{\per} = - \expo^{-\icomps t}\,\vecx\;\;
\mbox{for}\;\;|\vecr| \rightarrow \infty. \label{BC2bis} 
\end{align}
For notation convenience, we drop the primes in the rest of the paper.

\section{Inertial propulsion speed \label{inertial_effects}}

In the section, we consider the effects of inertia in the case of a near-sphere
oscillating in the transverse  direction in a prescribed way. We first expand
the velocity and pressure fields in powers of the Reynolds number. The perturbation in shape is introduced once
the governing equations are obtained at each order in Reynolds. We first consider in  \S\ref{zeroth_order} 
 the Stokes problem of an oscillating near-sphere (zeroth-order in Reynolds). 
In \S\ref{first_order} we introduce   the first-order (in Reynolds) problem. We then  use a suitable form of the reciprocal
theorem in \S\ref{prop_speed} in order to obtain the axial propulsion speed at leading order in an integral form, thereby
bypassing the calculation of the full flow at first order in Reynolds. The case of an arbitrary axisymmetric
shape is finally presented in  \ref{arbitrary_shape}.

We first expand the velocity,  pressure, and stress fields  in powers of the Reynolds number as follows
\begin{eqnarray}
\vecu & =& \vecu^{\sst (0)} + \Rey \,\vecu^{\sst (1)} + O(\Rey^2) \label{expand_u},\\[1mm]
p & =& p^{\sst (0)} + \Rey\,p^{\sst (1)} + O(\Rey^2)\label{expand_p},\\[1mm]
\vecs & =& \vecs^{\sst (0)} + \Rey \,\vecs^{\sst (1)} + O(\Rey^2). \label{expand_s}
\end{eqnarray}
The stress expansion is a consequence of the velocity and pressure expansions since at each order,
\begin{equation}
\vecs^{\sst (i)} = - p^{\sst (i)}\,\vecdelta + [\Nab \vecu^{\sst (i)} + \Nab {\vecu^{\sst (i)}}^{\dag}]\label{stress_order_i},
\end{equation}
where the superscript $\dag$ refers to the transposed tensor and $\vecdelta$ is the unit tensor.
Introducing equations (\ref{expand_u}) --(\ref{expand_s}) in the Navier-Stokes equations, 
equations (\ref{NS1_adim_2}) and (\ref{NS2_adim_2}), leads to the two sets of similar equations
satisfied by the zeroth and first order velocity/pressure fields. We consider them successively below.

\subsection{Zeroth-order solution in Reynolds \label{zeroth_order}}

The zeroth-order flow field satifies the Stokes equations
\begin{gather}
\lambda^2 \frac{\partial{\vecu^{\sst (0)}}}{\partial t} = -\Nab p^{\sst (0)} + \Nab^2\vecu^{\sst (0)},\label{NS01}\\
\Nab \cdot \vecu^{\sst (0)}  = 0,\label{NS02}
\end{gather}
with the  boundary conditions
\begin{align}
\vecu^{\sst (0)} & = \vecO\;\;\;\mbox{on $\mathcal{S}$},\label{BC01}\\
\vecu^{\sst (0)} & \rightarrow -\vecv^{\per} = -\expo^{-\icomps t}\,\vecx\;\;\mbox{for}\;\;|\vecr| \rightarrow \infty.\label{BC02}
\end{align}
Note that the oscillating transverse velocity is entirely taken into account in the
zeroth-order boundary conditions. Note also that no axial propulsion speed is expected at
that order since the kinematics corresponding to the transverse oscillation of the body can not lead to
any net force in the axial direction (reversibility). From here, we make the additional assumption that 
\begin{equation}
\lambda^2 \ll 1.
\end{equation}
This condition means that the viscous penetration   scale is much larger than the typical size $R_{\sst 0}$ of the
body. The flow is therefore approximately Stokesian in the entire space, enabling us to use Lorenz's reciprocal
theorem. In the opposite limit ($\lambda^2 \gg 1$), the viscous
flow would be  confined to a thin layer of thickness $\lambda^{-1}$, and the flow would be irrotational outside
the viscous layer \cite{Riley1966}.

In order to  obtain the right order in the final propulsion speed, we have
to expand the zeroth-order (in Reynolds) velocity and pressure fields to the first order in shape parameter,~$\eps$. We thus write
\begin{eqnarray}
\vecu^{\sst (0)} &=& \vecu^{\sst 0} + \eps\,\vecu^\eps + O(\eps^2),\label{expand_u_Re0_eps}\\
p^{\sst (0)} &=& p^{\sst 0} + \eps\,p^\eps + O(\eps^2), \label{expand_p_Re0_eps}
\end{eqnarray}
where $\vecu^{\sst 0}$ and $p^{\sst 0}$ are the velocity and pressure fields corresponding to the 
oscillations of the equivalent-volume sphere in a purely viscous fluid, and $\vecu^\eps$ and $p^\eps$ are the first 
corrections due to the difference in shape between the particle and the equivalent-volume sphere.

Working in Fourier space and denoting  Fourier transforms with a hat, 
the Fourier components of the velocity and pressure fields, $\hat{\vecu}^{\sst 0}$ and $\hat{p}^{\sst 0}$, satisfy the Stokes equations
\begin{gather}
- \icomp \lambda^2 \hat{\vecu}^{\sst 0} = -\Nab \hat{p}^{\sst 0} + \Nab^2\hat{\vecu}^{\sst 0},\label{NS01_bis}\\
\Nab \cdot \hat{\vecu}^{\sst 0}  = 0,\label{NS02_bis}
\end{gather}
together with the boundary conditions
\begin{align}
\hat{\vecu}^{\sst 0} & = \vecO\;\;\;\mbox{on $\mathcal{S}_{\sst 0}$},\label{osc_sph_BC1}\\
\hat{\vecu}^{\sst 0} & \rightarrow -\hat{\vecv}^{\per} = - \vecx\;\;\mbox{for}\;\;|\vecr| \rightarrow \infty.\label{osc_sph_BC2}
\end{align}
The Stokes flow induced by the oscillation of a sphere in a viscous fluid has been derived 
by Lamb \cite{Lamb} -- see also  \cite{Riley1966,Kim&Karrila}. This is a classical result and we recall its characteristics in  Appendix \ref{oscillating_sphere}. 

Due to the linear nature of the problem at zeroth-order in Reynolds, the corrective quantities
$\hat{\vecu}^\eps$ and $\hat{p}^\eps$ also satisfies the unsteady Stokes equations
\begin{gather}
- \icomp \lambda^2 \hat{\vecu}^\eps = -\Nab \hat{p}^\eps + \Nab^2\hat{\vecu}^\eps,\label{NS01_ter}\\
\Nab \cdot \hat{\vecu}^\eps  = 0.\label{NS02_ter}
\end{gather}
The fist boundary condition satisfied by the corrective flow $\hat{\vecu}^\eps$ is found by Taylor expanding
the boundary condition (\ref{BC01}) to the first order in $\eps$. Using equations (\ref{polar_adim})
and (\ref{expand_u_Re0_eps}), one then obtains the expression of the correction in shape on the
spherical surface $\mathcal{S}_{\sst 0}$ ({\it i.e.} at $r = 1$) as
\begin{equation}
\hat{\vecu}^\eps|_{r=\sst{1}} = -\xi(\theta)\,\left.\frac{\partial \hat{\vecu}^{\sst 0}}{\partial r}\right|_{r=\sst{1}}.
\end{equation}
As is recalled in Appendix \ref{oscillating_sphere}, the radial derivative of the  velocity field, 
$\hat{\vecu}^{\sst 0}$, is given  at the spherical surface $\mathcal{S}_{\sst 0}$ by
\begin{equation}
\left.\frac{\partial \hat{\vecu}^{\sst 0}}{\partial r}\right|_{r=\sst{1}} =
-\frac{3}{2}\,\hat{\vecv}_{\!\per}\cdot(1+\expo^{-\icomps\pi/4}\lambda)(\vecdelta - \vecn\vecn),
\end{equation}
where $\vecn$ is the unit vector normal to $\mathcal{S}_{\sst 0}$
which points towards the fluid (here $\vecn = \vecer$). Given that $\hat{\vecv}_{\!\per} = \vecx$, the explicit
form of the first boundary condition, expressed in the basis $(\vecer,\,\vectheta,\,\vecphi)$ of the spherical
coordinate system is thus
\begin{equation}
\hat{\vecu}^\eps|_{r=\sst{1}} = K\,\cos(n\theta)\,\left(
\begin{array}{c}
0\\
\cos\phi\cos\theta\\
-\sin\phi
\end{array}
\right),\label{stokes_BC1}
\end{equation}
with $K = (3/2)(1+\expo^{-\icomps\pi/4}\lambda)$.
As the corrective velocity flow (due to the difference in shape from that of the sphere) must vanish at large distances
from the particle, the following condition takes place
\begin{equation}
\hat{\vecu}^\eps \rightarrow \vecO\;\;\mbox{for}\;\;|\vecr| \rightarrow \infty. \label{BC_infty}
\end{equation}

The general form of the solution to the system formed by equations (\ref{NS01_ter}-\ref{NS02_ter}) has been derived by
Chandrasekhar \cite{Chandrasekhar} as a sum of spherical harmonics. 
Taking the curl of equation (\ref{NS01_ter}) leads to the equation which governs the vorticity.
After projecting the latter on the radial direction, one gets
\begin{equation}
(\icomp \lambda^2 + \Nab^2)(r\,\hat{\chi}^\eps) = 0,
\label{stokes_rad_vel}
\end{equation}
where $\hat{\chi}^\eps$ is the radial component of the vorticity. Similarly, an equation for the radial
component of the velocity $u^\eps_r$ is obtained by taking the radial component
of the curl of the vorticity equation and we obtain
\begin{equation}
\Nab^2(\icomp \lambda^2 + \Nab^2)(r\,\hat{u}_r^\eps) = 0.
\label{stokes_rad_vor}
\end{equation}
The objective is now to derive explicit expressions for  the radial components $\hat{u}_r^\eps$ and $\hat{\chi}^\eps$ of the
velocity and vorticity fields. In principle, the three components of the velocity must satisfy the boundary condition
(\ref{stokes_BC1}). Unfortunatly, only the radial components of velocity and vorticity are involved in the
governing equations (\ref{stokes_rad_vel}) and (\ref{stokes_rad_vor}). 
As 
is classically
done in such situations \cite{Miller1968}, we keep the  condition of continuity of
the radial component of the velocity
\begin{equation}
\hat{u}_r^\eps = 0\;\;\mbox{at}\;r = 1\label{rad_vel_BC},
\end{equation}
and  build two alternative boundary conditions
involving the surface divergence and the radial component of the surface curl of the velocity, by
recombining the velocity components (and their derivative) given by (\ref{stokes_BC1}).
These new boundary conditions are then used below instead of the continuity conditions on the polar and
azimuthal  velocity components. The advantage of such an approach is that $\partial_r \hat{u}_r^\eps|_{r=\sst{1}}$ 
and $\hat{\chi}^\eps$ are the only quantities involved in
the new set of boundary conditions. The surface divergence and the radial component of the surface curl of the
velocity at $r = 1$ are given by \cite{Kim&Karrila}
\begin{gather}
- r\,\Nab_s \cdot \hat{\vecu}^\eps = r\,\frac{\partial \hat{u}_r^\eps}{\partial r} = 
-2 \hat{u}_r^\eps - \frac{1}{\sin\theta}\frac{\partial}{\partial \theta}(\hat{u}_\theta^\eps\sin\theta)
-\frac{1}{\sin\theta}\frac{\partial \hat{u}_\phi^\eps}{\partial \phi},\label{surf_div}\\
r\,\vecer\cdot\Nab_s \times\hat{\vecu}^\eps = r\,\hat{\chi}^\eps = 
\frac{1}{\sin\theta}\frac{\partial}{\partial \theta}(\hat{u}_\phi^\eps\sin\theta)
-\frac{1}{\sin\theta}\frac{\partial \hat{u}_\theta^\eps}{\partial \phi},\label{surf_curl}
\end{gather}
where $\Nab_s = \Nab - \vecer \partial_r$ is the surface gradient operator.
After introducing the polar and azimutal components of $\hat{\vecu}^\eps$ at the surface 
$\mathcal{S}_{\sst 0}$ given by equations (\ref{stokes_BC1}) in   equations 
(\ref{surf_div}) and (\ref{surf_curl}), we obtain, for $r = 1$
\begin{gather}
- \Nab_s \cdot \hat{\vecu}^\eps = K \cos\phi\,[n\sin(n\theta)\cos\theta + 2 \cos(n\theta) \sin\theta],\label{surf_div_0}\\
\vecer\cdot\Nab_s\times\hat{\vecu}^\eps = K \sin\phi\,n \sin(n\theta).\label{surf_curl_0}
\end{gather}
We further show in Appendix \ref{transv_oscillation} that expressions (\ref{surf_div_0}) and (\ref{surf_curl_0}) 
of the surface divergence and curl can be rewritten as sums of associated Legendre functions of order 1. 
Thus, the previous equations can be put in the form
\begin{gather}
- \Nab_s \cdot \hat{\vecu}^\eps = K \cos\phi\,\sum_{q=0}^{k}B_{\sst{2}(q+1)} P^{\sst{1}}_{\sst{2}(q+1)}(\cos\theta) \label{surf_div_BC},\\
\vecer\cdot\Nab_s\times\hat{\vecu}^\eps = K \sin\phi\,\sum_{q=0}^{k}B_{\sst{2}q+1} P^{\sst{1}}_{\sst{2}q+1}(\cos\theta)\label{surf_curl_BC},
\end{gather}
where the constants $B_{\sst{2}q+1}$ and $B_{\sst{2}(q+1)}$ are also given in Appendix  \ref{transv_oscillation}.
Consequently, we can search for the radial components of velocity and vorticity in the form
\begin{align}
r\,\hat{u}_r^\eps = \sum_{q=0}^{k}r\,\hat{u}^\eps_{\sst{2}(q+1)},
\;\mbox{with}\;\;r\,\hat{u}^\eps_{\sst{2}(q+1)} = K\,U_{\sst{2}(q+1)}(r)\,P^{\sst{1}}_{\sst{2}(q+1)}(\cos\theta)\,\cos\phi, \label{sol_rad_vel}\\
r\,\hat{\chi}^\eps = \sum_{q=0}^{k}r\,\hat{\chi}^\eps_{\sst{2}q+1},
\;\mbox{with}\;\;r\,\hat{\chi}^\eps_{\sst{2}q+1} = K\,X_{\sst{2}q+1}(r)\,P^{\sst{1}}_{\sst{2}q+1}(\cos\theta)\,\sin\phi. \label{sol_rad_vor}
\end{align}
Introducing equations (\ref{sol_rad_vel}) and (\ref{sol_rad_vor}) into 
(\ref{stokes_rad_vel}) and (\ref{stokes_rad_vor}), and solving the resulting equations 
in $r$, one obtains the general forms of $U_{\sst{2}(q+1)}$ and $X_{\sst{2}q+1}$ as
\begin{align}
U_{\sst{2}(q+1)}&(r) =  \alpha_{\sst{2}(q+1)}^{\sst 0}  r^{2(q+1)}
+ \beta_{\sst{2}(q+1)}^{\sst 0}\left(\frac{\pi\expo^{-\icomps\pi/4}}{2\lambda r}\right)^{1/2}\,
J_{\sst{2}q+\frac{5}{2}}(\expo^{\icomps \pi/4}\lambda r) \nonumber\\
& + \alpha_{\sst{2}(q+1)}^{\sst \infty} r^{-(2q+3)} +
\beta_{\sst{2}(q+1)}^{\sst \infty}\left(\frac{\pi\expo^{-\icomps\pi/4}}{2\lambda r}\right)^{1/2}\,
H^{\sst (1)}_{\sst{2}q+\frac{5}{2}}(\expo^{\icomps \pi/4}\lambda r), \label{radial_factor_U}\\
X_{\sst{2}q+1}&(r) = \gamma_{\sst{2}q+1}^{\sst 0} \left(\frac{\pi\expo^{-\icomps\pi/4}}{2\lambda r}\right)^{1/2}\,
J_{\sst{2}q+\frac{3}{2}}(\expo^{\icomps \pi/4}\lambda r) \nonumber\\
&\hspace{2.5cm}+ \gamma_{\sst{2}q+1}^{\sst \infty} \left(\frac{\pi\expo^{-\icomps\pi/4}}{2\lambda r}\right)^{1/2}\,
H^{\sst (1)}_{\sst{2}q+\frac{3}{2}}(\expo^{\icomps \pi/4}\lambda r). \label{radial_factor_X}
\end{align}
In the previous expressions $J_l$ and $H^{\sst (1)}_l$ are Bessel functions and Hanckel
functions of the first kind respectively. Boundary condition (\ref{BC_infty}) imposes $\alpha_{\sst{2}(q+1)}^{\sst 0} = \beta_{\sst{2}(q+1)}^{\sst 0}
= \gamma_{\sst 2q+1}^{\sst 0} = 0$ allowing us to  drop the superscript $\infty$ in the following. The coefficients
$\alpha_{\sst{2}(q+1)}$, $\beta_{\sst{2}(q+1)}$ and $\gamma_{\sst 2q+1}$ are then to be determined using
the boundary conditions at the surface. After using equations (\ref{radial_factor_U}) and (\ref{radial_factor_X})
in the continuity conditions for the radial components of the velocity, equation (\ref{rad_vel_BC}), surface divergence, equation
(\ref{surf_div_BC}), and surface curl, equation (\ref{surf_curl_BC}), we obtain
\begin{gather}
\alpha_{\sst 2(q+1)} + \beta_{\sst 2(q+1)}\left(\frac{\pi\expo^{-\icomps\pi/4}}{2\lambda}\right)^{1/2}\,
H^{\sst (1)}_{\sst 2q+\frac{5}{2}}(\expo^{\icomps \pi/4}\lambda) = 0, \label{coef_radial_vel_BC}\\
-\alpha_{\sst 2(q+1)} (2q+4)+ \beta_{\sst 2(q+1)}\left(\frac{\pi\expo^{-\icomps\pi/4}}{2\lambda}\right)^{1/2}\,
\big[(2q+1)\,H^{\sst (1)}_{\sst 2q+\frac{5}{2}}(\expo^{\icomps \pi/4}\lambda) \hspace{2cm}\nonumber\\
 \hspace{5cm}-(\expo^{\icomps\pi/4}\lambda)\,H^{\sst (1)}_{\sst 2q+\frac{7}{2}}(\expo^{\icomps \pi/4}\lambda) 
\big] = B_{\sst 2(q+1)}, \label{coef_surf_div_BC}\\
\gamma_{\sst 2q+1} \left(\frac{\pi\expo^{-\icomps\pi/4}}{2\lambda}\right)^{1/2}\,
H^{\sst (1)}_{\sst 2q+\frac{3}{2}}(\expo^{\icomps \pi/4}\lambda) = B_{\sst 2q+1}.\label{coef_surf_curl_BC}
\end{gather}
When solved, the system of equations (\ref{coef_radial_vel_BC})--(\ref{coef_surf_div_BC}) gives the values
of $\alpha_{\sst 2(q+1)}$ and $\beta_{\sst 2(q+1)}$, while the last equation gives directly the value of $\gamma_{\sst 2q+1}$
\begin{eqnarray}
\alpha_{\sst 2(q+1)} &=& B_{\sst 2(q+1)}\left[\expo^{\icomps \pi/4}\lambda 
\frac{H^{\sst (1)}_{\sst 2q+\frac{7}{2}}(\expo^{\icomps \pi/4}\lambda)}{H^{\sst (1)}_{\sst 2q+\frac{5}{2}}(\expo^{\icomps \pi/4}\lambda)}-(4q+5)
\right]^{-1},\\
\beta_{\sst 2(q+1)} &=& -\sqrt{\frac{2}{\pi}} B_{\sst 2(q+1)} \left[
(\expo^{\icomps\pi/4}\lambda)^{1/2} H^{\sst (1)}_{\sst 2q+\frac{7}{2}}(\expo^{\icomps \pi/4}\lambda)
-(4q+5)(\expo^{\icomps\pi/4}\lambda)^{-1/2} H^{\sst (1)}_{\sst 2q+\frac{5}{2}}(\expo^{\icomps \pi/4}\lambda)\right]^{-1},\\
\gamma_{\sst 2q+1}  &=& B_{\sst 2q+1} \left[\left(\frac{\pi\expo^{-\icomps\pi/4}}{2\lambda}\right)^{1/2}\,
H^{\sst (1)}_{\sst 2q+\frac{3}{2}}(\expo^{\icomps \pi/4}\lambda)\right]^{-1}.
\end{eqnarray}
As shown in Ref.~\cite{Sani1963}, the complete velocity field $\hat{\vecu}^\eps$ can be reconstructed from the radial velocity
and vorticity components as
\begin{equation}
\hat{\vecu}^\eps = \hat{u}_r^\eps \vecer + \frac{r^2}{2(q+1)}
\left[\sum_{q=0}^{k}\frac{\Nab_s \mathcal{D} \hat{u}^\eps_{\sst 2(q+1)}}{(2q+3)}
 -\frac{\vecer \times \Nab_s\hat{\chi}^\eps_{\sst 2q+1}}{(2q+1)}\right],
\end{equation}
where the operator $\mathcal{D}$ is defined as
\begin{equation}
\mathcal{D}[...] = \frac{1}{r^2}\frac{\partial}{\partial r}\,\left[r^2...\right].
\end{equation}
The expressions we then obtain for  the components $\hat{u}^\eps_r$, $\hat{u}^\eps_\theta$, $\hat{u}^\eps_\phi$
of the flow field $\hat{\vecu}^\eps$ are given in Appendix \ref{transv_oscillation}.
 
\subsection{First-order solution \label{first_order}}

We now consider the derivation for the  first-order solution (in Reynolds) $\vecu^{\sst (1)}$. 
That flow field contains terms of different frequencies, but we are
here only interested in the steady part of the flow. For the sake of simplicity, we use $\vecu^{\sst (1)}$ to denote to the steady
component  of this first-order flow. The latter sastifies the following set of equations
\begin{gather}
\Nab\cdot \vecs^{\sst (1)} = \frac{1}{4}[(\ul{\hat{\vecu}}^{\sst (0)} \cdot \Nab)\,\hat{\vecu}^{\sst (0)}
+ (\hat{\vecu}^{\sst (0)} \cdot \Nab)\,\ul{\hat{\vecu}}^{\sst (0)}],\label{NS11}\\
\Nab \cdot \vecu^{\sst (1)}  = 0,\label{NS12}
\end{gather}
where  complex conjugate quantities are underlined. In the first-order governing
equations, the term $(\vecv^{\pa}\cdot\Nab) \vecu^{\sst (0)}$ has been dropped
since this term is
time-dependent (dimensionless frequency 1) and we are only interested in steady flows. 
Equations (\ref{NS11}) and (\ref{NS12}) have to be completed by the boundary conditions
\begin{align}
& \vecu^{\sst (1)} = \vecv^{\sst (1)}\;\;\;\mbox{on $\mathcal{S}$},\label{surf_vel_main}\\
& \vecu^{\sst (1)} \rightarrow \vecO \;\;\;\mbox{at infinity}.
\end{align}
where the unknown quantity  $\vecv^{\sst (1)}$ is linked to $\vecv^{\pa}$ by the relationship  
\begin{equation}
\vecv^{\pa} = \Rey\,\vecv^{\sst (1)}.
\label{prop_speed_adim}
\end{equation}

In order to obtain the first-order translation speed, we could try to derive the full velocity and stress fields
$\vecu^{\sst (1)}$ and $\vecs^{\sst (1)}$, and integrate the stress over the particle surface to obtain the propulsive force. However,
it is more convenient to use a suitable version of the reciprocal theorem, as suggested by Ho \& Leal
\cite{Ho1974} (the standard version of the Lorentz reciprocal  theorem can be found in Ref.~\cite{Kim&Karrila}).

\subsection{Reciprocal theorem and propulsion speed \label{prop_speed}}

For the same geometry, we consider now an auxiliary Stokes
velocity and stress fields $(\bar{\vecu},\,\bar{\vecs})$ satisfying 
\begin{gather}
\Nab\cdot\bar{\vecs} = \vecO,\label{NS_AUX1}\\
\Nab\cdot\bar{\vecu} = 0,\label{NS_AUX2}
\end{gather}
with suitable boundary conditions  to be specified below. 
Subtracting the inner product of equation (\ref{NS11}) with $\bar{\vecu}$ and the
inner product of equation (\ref{NS_AUX1}) with $\vecu^{\sst (1)}$, and integrating over 
the volume of fluid $\mathcal{V}$ leads to the equality of virtual powers as
\begin{align}
\int_{{\mathcal{V}}} [\bar{\vecu} \cdot (\Nab \cdot \vecs^{\sst (1)})\,- & ~\,\vecu^{\sst (1)}  
\cdot (\Nab\cdot\bar{\vecs})]\,d\mathcal{V} =  \nonumber \\ 
 & \frac{1}{4}\int_{{\mathcal{V}}} \bar{\vecu} \cdot [(\ul{\hat{\vecu}}^{\sst (0)} \cdot \Nab)\,\hat{\vecu}^{\sst (0)}
+ (\hat{\vecu}^{\sst (0)} \cdot \Nab)\,\ul{\hat{\vecu}}^{\sst (0)}]\,d\mathcal{V}. \label{RT_1}
\end{align}
Then, using the general vector identity
\begin{align}
\bar{\vecu} \cdot (\Nab \cdot \vecs^{\sst (1)})\,- & ~ \,\vecu^{\sst (1)} \cdot (\Nab\cdot \bar{\vecs}) = \nonumber \\ 
& \Nab \cdot (\bar{\vecu}\cdot \vecs^{\sst (1)} \,-\,\vecu^{\sst (1)} \cdot \bar{\vecs}) 
+ (\Nab \vecu^{\sst (1)} : \bar{\vecs} - \Nab \bar{\vecu} : \vecs^{\sst (1)}),\label{vect_identity} 
\end{align}
and realizing that the second term in the right-hand side  of equation (\ref{vect_identity}) vanishes for a Newtonian fluid,
we can rewrite equation (\ref{RT_1}) as
\begin{equation}
\int_{{\mathcal{V}}} \Nab \cdot (\bar{\vecu}\cdot \vecs^{\sst (1)} \,-\,\vecu^{\sst (1)} \cdot \bar{\vecs})\,d\mathcal{V} =
\frac{1}{4}\int_{{\mathcal{V}}} \bar{\vecu} \cdot [(\ul{\hat{\vecu}}^{\sst (0)} \cdot \Nab)\,\hat{\vecu}^{\sst (0)}
+ (\hat{\vecu}^{\sst (0)} \cdot \Nab)\,\ul{\hat{\vecu}}^{\sst (0)}]\,d\mathcal{V}. \label{RT_2}
\end{equation}
Using  the divergence theorem allows to simplify the left-hand side term and obtain
\begin{equation}
\int_{{\mathcal{S}}} \vecn \cdot (\bar{\vecu} \cdot \vecs^{\sst (1)}  - \vecu^{\sst (1)} \cdot \bar{\vecs}) \,d\mathcal{S} =
- \frac{1}{4}\int_{{\mathcal{V}}} \bar{\vecu} \cdot [(\ul{\hat{\vecu}}^{\sst (0)} \cdot \Nab)\,\hat{\vecu}^{\sst (0)}
+ (\hat{\vecu}^{\sst (0)} \cdot \Nab)\,\ul{\hat{\vecu}}^{\sst (0)}]\,d\mathcal{V}. \label{RT_3}
\end{equation}
We now define the boundary conditions for the auxiliary problem, $\bar{\vecu} $. We assume that it represents a solid-body motion 
with translational and angular velocities   $\bar{\vecv}$ and $\bar{\vecom}$, so that   
the auxiliary velocity at the surface $\mathcal{S}$ of the body is given by
\begin{equation}
\bar{\vecu} =  \bar{\vecv} + \bar{\vecom} \times \vecr, \label{surf_vel_aux}
\end{equation}
where $\vecr$ is the position vector. Since $\vecv^{\sst(1)}$ 
is the first-order propulsion speed, we can  introduce equations (\ref{surf_vel_main}) and (\ref{surf_vel_aux})
into equation (\ref{RT_3}), leading to the equality
\begin{align}
 \bar{\vecv} \cdot \! \int_{{\mathcal{S}}}  \vecn \cdot \vecs^{\sst (1)} \,d\mathcal{S}
+ &~ \bar{\vecom} \cdot \! \int_{{\mathcal{S}}} \vecr \times (\vecn \cdot  \vecs^{\sst (1)} )\,d\mathcal{S}
 - \vecv^{\sst(1)} \cdot \! \int_{{\mathcal{S}}} \vecn \cdot \bar{\vecs} \,d\mathcal{S} \nonumber\\
&  = -\frac{1}{4}\int_{{\mathcal{V}}} \bar{\vecu} \cdot [(\ul{\hat{\vecu}}^{\sst (0)} \cdot \Nab)\,\hat{\vecu}^{\sst (0)}
+ (\hat{\vecu}^{\sst (0)} \cdot \Nab)\,\ul{\hat{\vecu}}^{\sst (0)}]\,d\mathcal{V}.\label{RT_4} 
\end{align}
%
In  equation \eqref{RT_4}, the first term on the left-hand side is nothing but the inner product
of the auxiliary translational velocity of the solid body with the hydrodynamic force, 
$\vecf^{\sst(1)}$,  
in the main problem. The second term is the inner product of the auxiliary angular velocity
with the torque, $\vect^{\sst (1)}$, applied on the solid body by the main flow. The third term is of similar nature
to the first one with the role of the flows reversed.  Denoting by $\barf{\vecf}$ the force applied by the auxiliary
flow on the solid body, we obtain a convenient form of equation (\ref{RT_4}) as
\begin{equation}
\bar{\vecv} \cdot \vecf^{\sst(1)} + \bar{\vecom}\cdot \vect^{\sst (1)} - \vecv^{\sst(1)} \cdot \barf{\vecf}
= - \frac{1}{4}\int_{{\mathcal{V}}} \bar{\vecu} \cdot [(\ul{\hat{\vecu}}^{\sst (0)} \cdot \Nab)\,\hat{\vecu}^{\sst (0)}
+ (\hat{\vecu}^{\sst (0)} \cdot \Nab)\,\ul{\hat{\vecu}}^{\sst (0)}]\,d\mathcal{V}.\label{RT_5}  
\end{equation}

For the  problem of oscillation of the near sphere considered in this paper, only two quantities are important to 
compute in equation \eqref{RT_5}. Either the particle is free to move and we want to calculate $\vecv^{\sst(1)}$ 
or the particle is tethered and we wish to compute the hydrodynamics force applied by the fluid, balancing the 
external force tethering it.  In both cases, we can therefore pick  $\bar{\vecom}={\bf 0}$ and $\bar{\vecv}$ 
arbitrary. The flow with these boundary conditations has been calculated in the very general case of
an arbitrary near-sphere \cite{Happel&Brenner}. The particular case of an axisymmetric near-sphere in axial 
translation is presented in Appendix \ref{axial_translation}. As $\bar{\vecom} = \vecO$, equation (\ref{RT_5}) becomes
\begin{equation}
\vecv^{\sst(1)} \cdot \barf{\vecf} =\frac{1}{4} \int_{{\mathcal{V}}} \bar{\vecu} \cdot 
[(\ul{\hat{\vecu}}^{\sst (0)} \cdot \Nab)\,\hat{\vecu}^{\sst (0)}
+ (\hat{\vecu}^{\sst (0)} \cdot \Nab)\,\ul{\hat{\vecu}}^{\sst (0)}]\,d\mathcal{V}.\label{RT_swim}
\end{equation}
in the case of the force-free near-sphere ($\vecf^{\sst (1)}=\bf 0$), and  
\begin{equation}
\bar{\vecv} \cdot \vecf^{\sst(1)} = - \frac{1}{4}\int_{{\mathcal{V}}} \bar{\vecu} \cdot 
[(\ul{\hat{\vecu}}^{\sst (0)} \cdot \Nab)\,\hat{\vecu}^{\sst (0)}
+ (\hat{\vecu}^{\sst (0)} \cdot \Nab)\,\ul{\hat{\vecu}}^{\sst (0)}]\,d\mathcal{V}\label{RT_fixed}
\end{equation}
in the case of a tethered oscillating near-sphere ($\vecv^{\sst(1)} = \vecO$). 
Note that in these two equations, the magnitude of the solid body motion in the auxiliary problem is arbitrary,
since the hydrodynamic force scales linearly with the magnitude of the imposed velocity.

Focusing on the  force-free swimming case, we now consider the  expansion of  the right-hand side of  (\ref{RT_swim})  to order $O(\eps)$.
We first write the auxilliary flow as $\bar{\vecu} = \bar{\vecu}^{\sst 0} + \eps \bar{\vecu}^\eps$,
where $\bar{\vecu}^{\sst 0}$ is the field generated by an equivalent-volume sphere
translating at a velocity $\bar{\vecv}$ and $\bar{\vecu}^\eps$ the
perturbative field due to the non sphericity of the particle. We also expand the steady drag as 
$\barf{\vecf} = \barf{\vecf}^{\sst 0} +  \eps \barf{\vecf}^\eps$, where $\barf{\vecf}^{\sst 0}$ is the steady
drag of the sphere (of magnitude $-6\pi$) and $\barf{\vecf}^\eps$ the corrective drag due to the non sphericity
of the particle. The expressions of $\bar{\vecu}^{\sst 0}$ and
$\bar{\vecu}^\eps$ are both given in Appendix \ref{axial_translation}. Noticing that
\begin{equation}
\mathcal{V} = \mathcal{V}_{\sst 0} + \sum_i \mathcal{V}_{\sst +}^i + \sum_i \mathcal{V}_{\sst-}^i,
\end{equation}
where $\mathcal{V}_{\sst 0}$ is the volume of fluid outside the equivalent-volume sphere and
$\mathcal{V}_{\sst +}^i$ and $\mathcal{V}_{\sst-}^i$ are defined in figure \ref{geom},
and recalling that $\hat{\vecu}^{\sst (0)}$ can also be written as $\hat{\vecu}^{\sst 0} + \eps \hat{\vecu}^\eps$, equation 
(\ref{RT_swim}) can be expanded to order $O(\eps)$ to get formally
\begin{align}
\vecv^{\sst(1)} \cdot \left(\barf{\vecf}^{\sst 0} +  \eps \barf{\vecf}^\eps\right)& 
 = \frac{1}{4}\int_{{\mathcal{V}_{0}}} \bar{\vecu}^{\sst 0} \cdot [(\ul{\hat{\vecu}}^{\sst 0} \cdot \Nab)\,\hat{\vecu}^{\sst 0} 
+ (\hat{\vecu}^{\sst 0} \cdot \Nab)\,\ul{\hat{\vecu}}^{\sst 0}]\,d\mathcal{V} \nonumber\\
& - \frac{1}{4}\sum_i \int_{{\mathcal{V}_{\sst +}^i}} \bar{\vecu}^{\sst 0} \cdot [(\ul{\hat{\vecu}}^{\sst 0} \cdot \Nab)\,\hat{\vecu}^{\sst 0} 
+ (\hat{\vecu}^{\sst 0} \cdot \Nab)\,\ul{\hat{\vecu}}^{\sst 0}]\,d\mathcal{V}\nonumber\\
& + \frac{1}{4}\sum_i \int_{{\mathcal{V}_{\sst -}^i}} \bar{\vecu}^{\sst 0} \cdot [(\ul{\hat{\vecu}}^{\sst 0} \cdot \Nab)\,\hat{\vecu}^{\sst 0} 
+ (\hat{\vecu}^{\sst 0} \cdot \Nab)\,\ul{\hat{\vecu}}^{\sst 0}]\,d\mathcal{V}\nonumber\\
& + \,\frac{\eps}{4}\, \bigg[
\int_{{\mathcal{V}}} \bar{\vecu}^\eps \cdot [(\ul{\hat{\vecu}}^{\sst 0} \cdot \Nab)\,\hat{\vecu}^{\sst 0} 
+ (\hat{\vecu}^{\sst 0} \cdot \Nab)\,\ul{\hat{\vecu}}^{\sst 0}]\,d\mathcal{V}\nonumber\\
+ \int_{{\mathcal{V}}} \bar{\vecu}^{\sst 0} \cdot [(\ul{\hat{\vecu}}^{\sst 0} \cdot \Nab)&\,\hat{\vecu}^\eps 
+ (\hat{\vecu}^{\sst 0} \cdot \Nab)\,\ul{\hat{\vecu}}^\eps + (\ul{\hat{\vecu}}^\eps \cdot \Nab)\,\hat{\vecu}^{\sst 0} 
+ (\hat{\vecu}^\eps \cdot \Nab)\,\ul{\hat{\vecu}}^{\sst 0}]\,d\mathcal{V} \bigg]
\label{RT_6}
\end{align}
The first term of the right-hand side of  (\ref{RT_6}) vanishes  since it corresponds
to the translational speed of a sphere oscillating in a viscous fluid (which is zero by symmetry \cite{Riley1966}). Furthermore, since 
the particle is nearly spherical, volume integrals in the second and third terms can be replaced by surface integrals for
\begin{align}\label{member}
\int_{{\mathcal{V}_{\sst \pm}^i}} \bar{\vecu}^{\sst 0} \cdot [(\ul{\hat{\vecu}}^{\sst 0}& \cdot \Nab)\,\hat{\vecu}^{\sst 0} 
+ (\hat{\vecu}^{\sst 0} \cdot \Nab)\,\ul{\hat{\vecu}}^{\sst 0}]\,d\mathcal{V} = \nonumber\\
& \eps\,\int_{{\mathcal{S}_{\sst \pm}^i}} \xi\,\big[\bar{\vecu}^{\sst 0} \cdot [(\ul{\hat{\vecu}}^{\sst 0} \cdot \Nab)\,\hat{\vecu}^{\sst 0} 
+ (\hat{\vecu}^{\sst 0} \cdot \Nab)\,\ul{\hat{\vecu}}^{\sst 0}]\big]_{{\mathcal{S}_{\sst 0}}}\,d\mathcal{S} + O(\eps^2),
\end{align}
where the surfaces $\mathcal{S}_{\sst +}^i$ and $\mathcal{S}_{\sst -}^i$ are defined in figure \ref{geom}.
Since $\hat{\vecu}^{\sst 0}$ is identically zero on the unit sphere, the integral on the right in 
equation \eqref{member} is zero. Consequently, the second and third terms of the right-hand side of equation (\ref{RT_6}) can also be neglected
at order $O(\eps)$. Furthermore, integrating the last two terms of the right-hand side of (\ref{RT_6})
over $\mathcal{V}_{\sst 0}$ instead of $\mathcal{V}$ induces no change at order $O(\eps)$ since
the error introduced by doing this is of order $O(\eps^2)$. As a  consequence, the corrective drag, $\barf{\vecf}^\eps$, does not need
to be computed  since it only involves correction of  $O(\eps^2)$
in the final result. Finally, one then obtains an order $\epsilon$ swimming velocity
\begin{equation}
\vecv^{\sst (1)} = \eps\,\vecv^{\sst(1,1)} ,
\end{equation}
where
\begin{align}
\vecv^{\sst(1,1)}  = \,&\frac{1}{4}\,\big|\barf{\vecf}^{\sst 0}\big|^{-1}\,\vecz\, \bigg[
\int_{{\mathcal{V}_{\sst 0}}} \bar{\vecu}^\eps \cdot [(\ul{\hat{\vecu}}^{\sst 0} \cdot \Nab)\,\hat{\vecu}^{\sst 0} 
+ (\hat{\vecu}^{\sst 0} \cdot \Nab)\,\ul{\hat{\vecu}}^{\sst 0}]\,d\mathcal{V}\nonumber\\
+ \int_{{\mathcal{V}_{\sst 0}}} & \bar{\vecu}^{\sst 0} \cdot [(\ul{\hat{\vecu}}^{\sst 0} \cdot \Nab)\,\hat{\vecu}^\eps 
+ (\hat{\vecu}^{\sst 0} \cdot \Nab)\,\ul{\hat{\vecu}}^\eps + (\ul{\hat{\vecu}}^\eps \cdot \Nab)\,\hat{\vecu}^{\sst 0} 
+ (\hat{\vecu}^\eps \cdot \Nab)\,\ul{\hat{\vecu}}^{\sst 0}]\,d\mathcal{V} \bigg]
\label{expr_prop_speed}
\end{align}
and is an $O(1)$ quantity. The superscript $(1,1)$ is used to remind us that the leading-order dimensionless propulsion speed, 
$\vecv^{\pa}$, scales as the first power of the shape parameter and the Reynolds number.

The expressions for the fields $\hat{\vecu}^{\sst 0}$, $\hat{\vecu}^\eps$ and $\bar{\vecu}^\eps$  necessary to compute equation \eqref{expr_prop_speed}
are given in appendices \ref{oscillating_sphere}, \ref{transv_oscillation}, and \ref{axial_translation} respectively. With these solutions knowns, the
gradients $\Nab \hat{\vecu}^\eps$ and $\Nab \hat{\vecu}^{\sst 0}$ can be formally computed (their lengthy expressions are not reproduced here to spare the reader). 
In the following, the quantity $\vecv^{\sst (1,1)}$ given by equation (\ref{expr_prop_speed}) will be denoted by $\vecv_{k}^{\sst (1,1)}$, with 
the subscript $k$ simply used to remind that this expression has been derived for a shape function of
the form $\cos(n\theta)$ with $n = 2k+1$. Once $\vecv_{k}^{\sst (1,1)}$ is known, the dimensional velocity $\vecV^{\pa}_k$
for the mode $k$ can be deduced immediately as
\begin{equation}
\vecV^{\pa}_k = \eps\Rey\,\hat{V}^{\per}\,\vecv_{k}^{\sst (1,1)}.
\end{equation}
As a concluding note, we point out that the propulsion speed
does not depend on the choice made for the origin of the coordinate system, as demonstrated in  Appendix \ref{origin_position}.  Similarly, the
propulsion speed does not depend on the precise definition of  $R_{\sst 0}$ either,
since a change of $R_{\sst 0}$ of order $O(\eps R_{\sst 0})$ would also lead to corrections of order $O(\eps^2)$ in equation (\ref{expr_prop_speed}). 

\subsection{Case of an arbitrary axisymmetric shape \label{arbitrary_shape}}

We now consider the case of a particle of arbitrary axisymmetric shape. The polar equation of the particle is still
given by equation (\ref{polar}) where $\xi$ is  an arbitrary order one function of $\theta$ defined on the interval $[0,\,\pi]$.
The Fourier-cosine series for this function can be written down as
\begin{equation}
\xi(\theta) = \sum_{n=0}^{\infty} \zeta_n \cos(n\theta).
\label{shape_Fourier}
\end{equation}
Equation (\ref{expr_prop_speed}) denotes the propulsion speed, $\vecv_{k}^{\sst (1,1)}$,   for a  shape function of the form $\cos(n\theta)$ 
with $n = 2k+1$, and   the propulsion speed is exactly zero for 
even values of $n$ by symmetry. Since the  perturbed fields $\bar{\vecu}^\eps$ and $\vecu^\eps$ have no quadratic contribution in the
integrands involved in (\ref{expr_prop_speed}), we can write the propulsion speed for any axisymmetric arbitrary shape
as a linear superposition 
\begin{equation}
\vecv^{\sst (1,1)} = \sum_{k=1}^{\infty} \zeta_{\sst 2k+1}\,\vecv_{k}^{\sst (1,1)},
\end{equation}
allowing to  compute the dimensional propulsion speed
\begin{equation}
\vecV^{\pa} = \eps\Rey\,\hat{V}^{\per}\,\vecv^{\sst (1,1)} 
\label{dimensional_prop_speed},
\end{equation}
of any axisymmetric arbitrary near-sphere oscillating in the transverse direction once the $\vecv_{k}^{\sst (1,1)}$ and the Fourier coefficients of $\xi$
are known. By replacing the typical velocity $\hat{V}^{\per}$ and the Reynolds number
by their expression in terms of physical parameters of the problem, we obtain the final  form of the propulsion speed as
\begin{equation}
\vecV^{\pa} = \eps\,\frac{a^2\omega^2\,R_{\sst 0}}{\nu}\,\vecv^{\sst(1,1)}.
\end{equation}

\section{Computation of the propulsion speed \label{num_int}}

\begin{figure}[t]
\scalebox{0.9}{\includegraphics{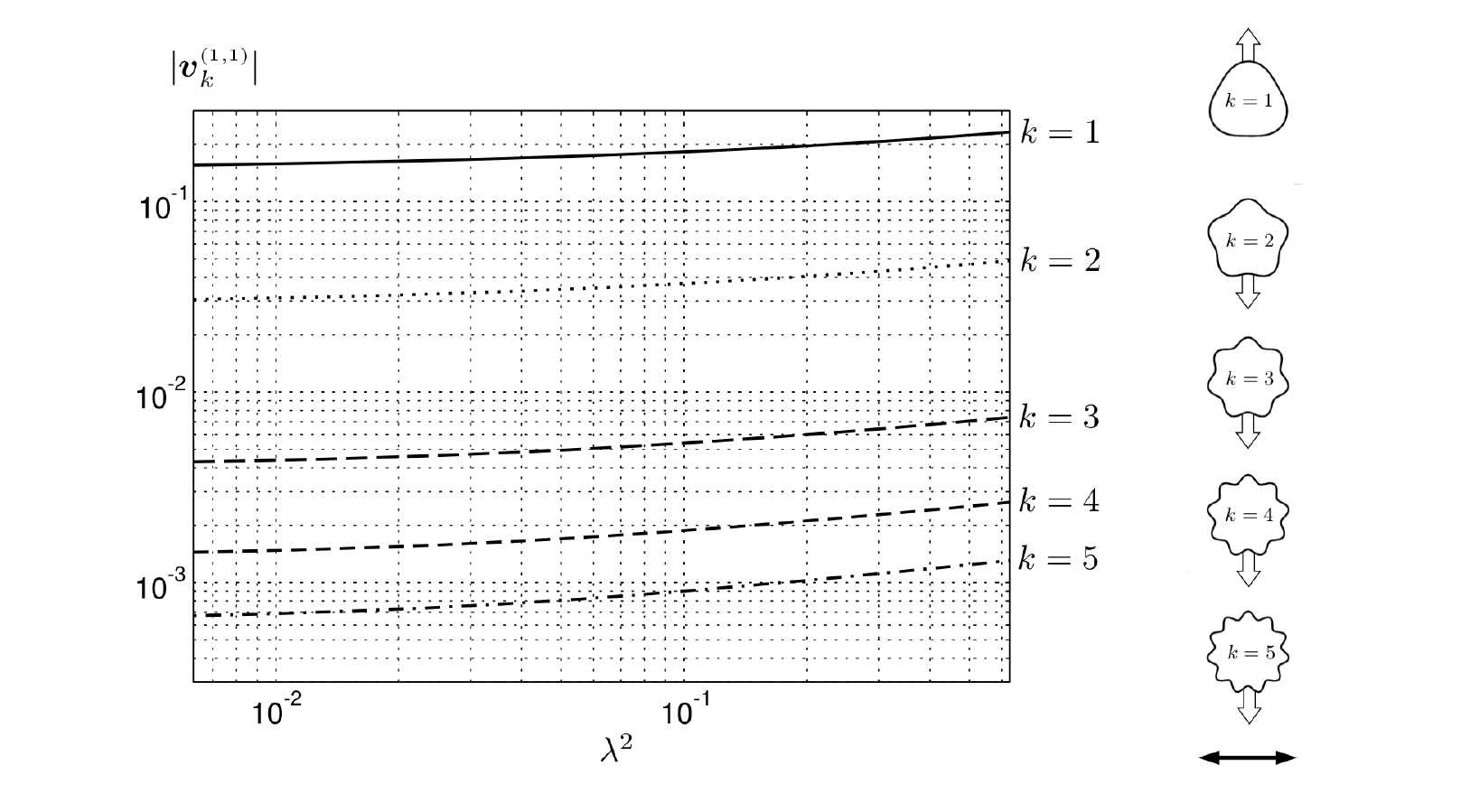}}
\caption{Velocity magnitude,  $|\vecv_{k}^{\sst (1,1)}|$, as a function of the parameter $\lambda^2$ for the first five modes ($k=1\to 5$) in 
the range $\lambda^2 \in [6.28\times10^{-3},\,6.28\times10^{-1}]$. The propulsion speed is oriented along $+z$ for the mode $k = 1$ and along 
$-z$ for the other four modes. The previous observation is schematized on the right side of the main graph. The double black arrow
stands for the direction of oscillation, that is the direction of the wavenumber in the case of a particle trapped
at the pressure node of an acoustic standing wave. For each shape, the direction of propulsion is given by a white arrow.
}
\label{dimless_prop_speed}
\end{figure}

In order to enable the calculation of the propulsion speed in the case of arbitrary shapes, the quantity $\vecv_{k}^{\sst (1,1)}$ has
been computed numerically for the first five modes ($k=1$ to $5$).  The numerical integration of equation  (\ref{expr_prop_speed}) has 
been performed in the  $(r,\theta,\phi)$ space using Matlab. The space occupied by the fluid corresponds to the interval 
$[1,\infty]\times[0,\pi]\times[0,2\pi]$, but we limited the integration in the radial direction
to the value $r = 10$. We discretized the range $[1,10]$, $[0,\pi]$, and $[0,2\pi]$ into  500, 180 and 360 intervals respectively. 
The computation  was done by means of  Legendre-Gauss Quadratures. We computed the results   for values of the dimensionless frequency 
parameter $\lambda^2$ in the range $[6.28\times10^{-3},\,6.28\times10^{-1}]$,  corresponding, for a particle of size 
$R_{\sst 0} = 1\,\mu$m in water, to the relevant  range of frequency of 1 to 100~kHz. 
In figure \ref{dimless_prop_speed}, we plot the velocity magnitude,  $|\vecv_{k}^{\sst (1,1)}|$,  as a function of the dimensionless parameter, $\lambda^2$.
The quantity  $|\vecv_{k}^{\sst (1,1)}|$ is a decreasing function of $k$ and a slowly increasing function of $\lambda^2$ on the considered interval.
The direction of propulsion (oriented along the $z$-axis) is observed to be  positive for $k = 1$ and negative for $k = 2,\,3,\,4$ and 5.
As an example, if we consider the case of a nearly spherical particle with $k = 1$, $\eps = 0.1$,  with radius $R_{\sst 0} = 1\,\mu$m,
oscillating in water with a pulsation $\omega = 100$ kHz and an amplitude  $a = 0.1\,\mu$m, we obtain numerically $V^{\pa}_{\sst 1} = 0.456\,\mu$m s$^{-1}$.

\section{Near-sphere in a uniform oscillating velocity field \label{acoustic_field}}

In the previous sections, we  considered the axial motion of an axisymmetric body in a quiescent fluid. The body was assumed to be force-free in
the axial $z$-direction and its transverse harmonic motion (along $x$) was fully prescribed.  Here we turn to the  problem of the dynamic response
of the same axisymmetric body in a uniform  oscillating exterior velocity field, $\vecU_e = \hat{\vecU}_e\,\expo^{-\icomps \omega T}$. This situation
occurs for instance after a solid particle drifted and is trapped at a pressure node of a standing sound wave \cite{Doinikov1994a}. Specifically, in the low
frequencies regime, $\lambda^2 \ll 1$, such drifting occurs for  particles of density $\rho_p$ less than twice the fluid density ($\rho_p<2\rho$),
and  for  particles of density $\rho_p>(2/5)\rho$ in the high frequency limit,  $\lambda^2 \gg 1$. We assume that  the axis of symmetry of the body is on average 
perpendicular to the external flow direction, $\hat{\vecU}_e$ and if this was not the case, hydrodynamic torques would rotate the particle into that configuration 
by symmetry. 

One important point needs to be noted here. Considering the the particle as oscillating in a locally uniform oscillatory velocity field is not a good approximation  if it is located at an arbitrary position in the acoustic field. Specifically, if the particle is far from a pressure node (velocity loop), the surrounding incident velocity field contains a linear component leading to a dipolar streaming flow parallel to the wave vector \cite{Danilov2000,Gorkov1962}. This is further discussed in  Appendix \ref{dipolar_quadrupolar}. The calculations in our paper focus on the dynamics after the particle has been trapped at the pressure node.

In order to compute the swimming speed, we first have to characterize the unsteady Stokes problem ($\Rey = 0$) with the  mass of the body now taken into account. The oscillating particle experiences time-dependent  hydrodynamics  torques or forces. Consequently, it will not only translate along the transverse $x$-direction, but will also rotate around the $y$-direction, and  the body is now neither torque-free in the $y$-direction nor force-free in the transverse $x$-direction since its own inertia is not neglected.

We consider again  an homogeneous solid particle (density $\rho_p$) 
the shape of which is again defined by its polar equation $R = R_{\sst 0}[1 + \eps\,\xi(\theta)]$. The radius $R_{\sst 0}$ and the  position of the origin of the coordinate system are  however chosen so that equations (\ref{cond_1}) and (\ref{cond_2}) are satisfied. This  means that the volume of the  near sphere is  $\mathcal{V}_p = (4/3) \pi R_{\sst 0}^3$ and  the origin of the coordinate system coincides with the center of gravity of the particle.  This choice does not modify equation  (\ref{expr_prop_speed}) derived in the previous section (Appendix \ref{origin_position}) and allows a convenient application of the theorem of angular momentum free from additional inertial terms.

\subsection{Prescribed rotational and translational motion in a quescient fluid \label{prescribed_motion}}

We first detail the case in which the transverse translational and rotational motions of the body are prescribed.  A 
particle translating with velocity $\vecV = \hat{\vecV}\,\expo^{-\icomps \omega T}$ and rotating with
angular velocity $\vecOm = \hat{\vecOm}\,\expo^{-\icomps \omega T}$ in a quiescent fluid will experience a hydrodynamic
force $\vecF = \hat{\vecF}\,\expo^{-\icomps \omega T}$ and a hydrodynamic torque 
$\vecL = \hat{\vecL}\,\expo^{-\icomps \omega T}$ where $\hat{\vecF}$ and $\hat{\vecL}$  are linearly related to $\hat{\vecV}$
and $\hat{\vecOm}$ as \cite{Zhang1998}
\begin{equation}
\left(
\begin{array}{c}
\hat{\vecF}\\
\hat{\vecL}/R_{\sst 0}
\end{array}
\right)
=
-6\pi\mu R_{\sst 0}\,\left(
\begin{array}{cc}
\dsty \vecA & \vecB \\
\dsty \vecB^{\sst T} & \frac{4}{3}\vecC
\end{array}
\right)
\cdot
\left(
\begin{array}{c}
\hat{\vecV}\\
R_{\sst 0} \hat{\vecOm}
\end{array}
\right),
\label{coupling_gene}
\end{equation}
where the tensors $\vecA$ and $\vecC$ can be expanded to order $O(\eps^2)$ in the form 
\begin{gather}
\vecA = \Lambda_{\sst 0}\,\vecdelta + \eps\,\vecA^{\eps} + \eps^2 \vecA^{\eps\eps} + O(\eps^3),\\
\vecC = \Delta_{\sst 0}\,\vecdelta + \eps\,\vecC^\eps + \eps^2 \vecC^{\eps\eps} + O(\eps^3).
\end{gather}
The $O(1)$ and $O(\eps)$ terms  of these tensors have been calculated analytically   by Zhang \& Stone \cite{Zhang1998}  using
  an appropriate version of the reciprocal theorem. They obtained explicitly 
\begin{gather}
\Lambda_{\sst 0} = 1 + \expo^{-\icomps \pi/4}\lambda - \icomp\frac{\lambda^2}{9},\label{Lambda0}\\
\Delta_{\sst 0} = \frac{1}{3}\left(\frac{\icomp\lambda^2}{\expo^{-\icomps \pi/4}\lambda +1} + 3\right),\label{Delta0}
\end{gather}
and
\begin{gather}
\vecA^\eps = \frac{9}{24\pi}(\expo^{-\icomps \pi/4}\lambda +1)\,\int_{\mathcal{S}}\xi\,\vecn\vecn\,d\mathcal{S},\\
\vecC^\eps = \frac{1}{8\pi}\left[\left(\frac{\icomp\lambda^2}{\expo^{-\icomps \pi/4}\lambda +1} + 3\right)^2
-\icomp\lambda^2\right]\,\int_{\mathcal{S}}\xi\,\vecn\vecn\,d\mathcal{S},
\end{gather}
Furthermore, provided that the torque is calculated at the center of mass (so that equation (\ref{cond_2}) is satisfied),
the coupling tensor $\vecB$ is of order $O(\eps^2)$ at most \cite{Zhang1998}. Since by  symmetry, an
axisymmetric body  oscillating in a transverse direction ($X$-direction) relative to its axis of symmetry ($Z$-direction),
does not experience any axial oscillating force,   equation (\ref{coupling_gene}) can be written in the $(\vecx,\,\vecy)$
basis in the simpler form
\begin{equation}
\left(
\begin{array}{c}
\hat{F}\\
\hat{L}/R_{\sst 0}
\end{array}
\right)
=
- 6\pi\mu R_{\sst 0}\,\left(
\begin{array}{cc}
\dsty A & \eps^2\,B^{\eps\eps}\\
\dsty \eps^2 \,B^{\eps\eps}  & \frac{4}{3} C
\end{array}
\right)
\cdot
\left(
\begin{array}{c}
\hat{V}\\
R_{\sst 0} \hat{\Omega}
\end{array}
\right)+ O(\eps^3)
\label{coupling_gene_2},
\end{equation}
where all terms, namely $\hat{F}$, $\hat{L}$, $\hat{V}$, $\hat{\Omega}$, 
$A = \Lambda_{\sst 0} + \eps\,A^{\eps} + \eps^2 A^{\eps\eps} + O(\eps^3)$, 
$B^{\eps\eps}$, $C = \Delta_{\sst 0} + \eps\,C^{\eps} + \eps^2 C^{\eps\eps} + O(\eps^3)$,  
are now scalar quantities.

\subsection{Dynamic response at zero Reynolds number\label{dynamic_response_af}}

We now address the main problem of the dynamic response of the particle in the uniform oscillating exterior field, 
$\vecU_e = \hat{U}_e\,\expo^{-\icomps \omega T}\,\vecx$. Under the effect of the
exterior field, the particle oscillates in the $x$-direction with velocity 
$\vecV = \hat{V}\,\expo^{-\icomps \omega T}\,\vecx$ in the laboratory frame and rotates about the $y$-axis
with angular velocity $\vecOm = \hat{\Omega}\,\expo^{-\icomps \omega T}\,\vecy$.
In order to derive the dynamics for the particle at leading order, we begin by
 considering the force experienced by a particle oscillating in a uniform oscillating
Stokes flow field. Working in the reference frame of the oscillating fluid (the reference frame
in which the fluid is motionless at large distances from the particle) the perturbed flow resulting from the
presence of the particle is governed by the unsteady Stokes equations written in the Fourier space as
\begin{gather}
-\icomp \omega \hat{\vecU} = -\frac{1}{\rho} \Nab \hat{P} + \nu\Delta \hat{\vecU},\\
\Nab\cdot \hat{\vecU} = 0,
\end{gather}
where $\hat{\vecU}$ and $\hat{P}$ are the Fourier component of the velocity and pressure fields.
Note that the inertial force density due to the acceleration of the reference frame,  $\rho \hat{\vecGamma}_e = \icomp \rho\,\omega \hat{U}_e\vecx$, 
 can be incorporated in the pressure gradient term, since it can be written as
minus the gradient of the pressure $\hat{P}_e = - \icomp \rho\,\omega \hat{U}_e X$. The boundary conditions satisfied by the flow field $\hat{\vecU}$ are
\begin{align}
& \hat{\vecU} = (\hat{V} - \hat{U}_e)\,\vecx\;\mbox{on $\mathcal{S}$},\\
& \hat{\vecU} = \vecO\;\;\mbox{for}\; \vecr \rightarrow \infty.
\end{align}
The problem is formally the same as that of a particle oscillating with velocity $(\hat{V} - \hat{U}_e)\,\vecx$
in a quiescent fluid considered above. The integration over the particle surface of the stress tensor corresponding
to the fields $\hat{\vecU}$ and $\hat{P}$  leads then to expressions for the force and torques $\hat{\vecF}' = \hat{F}'\,\vecx$ and $\hat{\vecL}' = \hat{L}'\,\vecy$ given by
\begin{equation}
\left(
\begin{array}{c}
\hat{F}'\\
\hat{L}'/R_{\sst 0}
\end{array}
\right)
=
-6\pi\mu R_{\sst 0}\,\left(
\begin{array}{cc}
A & \eps^2 B^{\eps\eps} \\
\eps^2 B^{\eps\eps} &  \frac{4}{3}C
\end{array}
\right)
\cdot
\left(
\begin{array}{c}
\hat{V} - \hat{U}_e\\
R_{\sst 0} \hat{\Omega}
\end{array}
\right)+ O(\eps^3)
\label{coupling_gene_3},
\end{equation}
In the present situation, this force has little physical meaning since the frame of reference we work in is not Galilean.
To obtain the expression of the actual force experienced by the particle (which should, of course, not
be depending on the reference frame), we must subtract the effect of the inertial pressure
$\hat{P}_e$. Integrating the latter over the particle surface leads to an additional force
\begin{equation}
\hat{\vecF}_e  = \int_{\mathcal{S}}\hat{P}_e\,\vecn\,d\mathcal{S} 
= \hat{F}_e\vecx = -\icomp\,\mathcal\rho\,\omega\hat{U}_e\,\mathcal{V}_{\sst p}\,\vecx,
\end{equation}
but there is no contribution to the torque\footnote{The contribution to the torque, $\hat{\vecL}_e$,  calculated at the center of
gravity of an homogeneous body  with uniform pressure gradient is zero. We have
\begin{equation}
\hat{\vecL}_e = \int_{\mathcal{S}} \vecr \times \hat{P}_e\vecn\,d\mathcal{S} 
= - \int_{\mathcal{V}_p} \Nab \times \hat{P}_e\vecr\,d\mathcal{V},
\hat{\vecL}_e = - \Nab\hat{P}_e \times \int_{\mathcal{V}_p}\vecr\,d\mathcal{V}
- \int_{\mathcal{V}_p} \hat{P}_e\,\Nab \times \vecr\,d\mathcal{V}.
\end{equation}
The first term on the right-hand side of this expression vanishes as the origin of the coordinate
system is located at the centre of gravity of the solid body and the second one is  zero as $ \Nab \times \vecr= \vecO$.}.
The final expression for the total hydrodynamic force experienced by the particle is given by
\begin{gather}
\left(
\begin{array}{c}
\hat{F}\\
\hat{L}/R_{\sst 0}
\end{array}
\right)
=
-6\pi\mu R_{\sst 0}\,\left(
\begin{array}{cc}
A & \eps^2 B^{\eps\eps} \\
\eps^2 B^{\eps\eps} &  \frac{4}{3}C
\end{array}
\right)
\cdot
\left(
\begin{array}{c}
\hat{V} - \hat{U}_e\\
R_{\sst 0} \hat{\Omega}
\end{array}
\right)
+
\left(
\begin{array}{c}
\hat{F}_e\\
0
\end{array}
\right)+ O(\eps^3)
\label{coupling_unif_field}.
\end{gather}

We now apply the momentum theorems (translational and angular) to the particle, keeping in mind that the angular momentum theorem
applied at the center of gravity of a solid body   has the same form  in any Galilean reference frame. The governing equations for
the dynamics of the particle are then given, at order $O(\eps^2)$, by 
\begin{eqnarray}
-\icomp\rho_p\mathcal{V}_p \hat{V}& =& - 6\pi\mu R_{\sst 0}\,A\,(\hat{V} - \hat{U}_e) 
+ \hat{F}_e - \eps^2\,6\pi\mu R_{\sst 0}^2 B^{\eps\eps}\,\hat{\Omega},\\
-\icomp \mathcal{I}_p\,\hat{\Omega} &=& - \eps^2\,6\pi\mu R_{\sst 0}^2 B^{\eps\eps}\,(\hat{V} - \hat{U}_e)
- 8\pi\mu\,R_{\sst 0}^3C\,\hat{\Omega},
\label{coupling_unif_field_2} 
\end{eqnarray}
where the moment of inertia of the particle,  $\mathcal{I}_p$, can be written to order $O(\eps^2)$ as
\begin{equation}
\mathcal{I}_p = \mathcal{I}_{\sst 0} (1+\eps\eta^\eps +\eps^2\eta^{\eps\eps}) + O(\eps^3),
\end{equation}
with $\mathcal{I}_{\sst 0} = (8/15)\pi\rho_p\,R_{\sst 0}^5$  the moment of inertia of the equivalent-volume sphere
about the $y$-axis. From these equations, we then obtain $\hat{\Omega}$ and $\hat{V}$ at leading order in $\eps$ 
\begin{gather}
\hat{V} = \frac{9 \beta \icomp \,\Lambda_e}{2\lambda^2 + 9 \icomp \beta \Lambda_{\sst 0}}\,\hat{U}_e + O(\eps),\label{V_coupl}\\[2mm]
R_{\sst 0} \hat{\Omega} = \icomp \eps^2\,
\frac{90\beta B^{\eps\eps} \lambda^2(1-\beta)}{(2\lambda^2+9\icomp\beta \Lambda_{\sst 0})
(4\lambda^2 + 60\icomp\beta \Delta_0)}\,\hat{U}_e + O(\eps^3),\label{omega_coupl}
\end{gather}
where
\begin{equation}
\Lambda_e = 1 + \expo^{-\icomps \pi/4}\lambda - \icomp\frac{\lambda^2}{3},
\end{equation}
 $\beta = \rho/\rho_p$ and $\Lambda_{\sst 0}$ and $\Delta_{\sst 0}$
are given by equations (\ref{Lambda0}) and (\ref{Delta0}) respectively. 
From equation  (\ref{omega_coupl}), we can deduce that the rotational motion leads to a Stokes flow of order $\eps^2$ and consequently,
does not change the analysis presented in sections \ref{presentation} and \ref{inertial_effects}. Finally, the relative amplitude of the
particle oscillations is obtained as
\begin{equation}
\hat{\vecV}^{\per} =  \hat{\vecV} - \hat{\vecU}_e = (\beta-1)\,\frac{2\lambda^2}{2\lambda^2 + 9 \icomp \beta \Lambda_{\sst 0}}\,\hat{U}_e\,\vecx.
\label{rel_transv_speed}
\end{equation}

A particle taking the shape of  an axisymmetric near-sphere whose transverse motion is forced by an oscillating 
uniform velocity field is thus propelled  with   dimensional  swimming  speed given  by  equation (\ref{dimensional_prop_speed}),
with a  Reynolds number in which $\hat{V}^{\per}$ is given by the norm of equation (\ref{rel_transv_speed}). Note that from equation
(\ref{rel_transv_speed}), we observe that the density of the particle must be different from the density of the surrounding fluid
($\beta \ne 1$) for the relative velocity, and consequently the Reynolds number, to be non-zero.

\section{Discussion}
\label{disc_concl}

In this paper we presented  a mechanism of propulsion for solid particles based on steady streaming. We show how the transverse oscillations
of an asymmetric shape gives rise, in general, to a non-zero time average propulsive force in the direction perpendicular to that of the imposed
oscillations. The calculations were made under the assumption of near-sphericity ($\eps\ll 1$) and small Reynolds number ($\Rey\ll 1$),
leading to a free-swimming speed of a particle, $V^{\pa}$,  scaling as
\begin{equation}
V^{\pa} = \eps\Rey\,\hat{V}^{\per}\,v^{\sst(1,1)},
\end{equation}
where $\hat{V}^{\per}$ denotes the amplitude of the transverse oscillations and $v^{\sst(1,1)}$ is of order one and given by equation (\ref{expr_prop_speed}).

\begin{figure}
\hspace*{0cm}
\scalebox{0.9}{\includegraphics{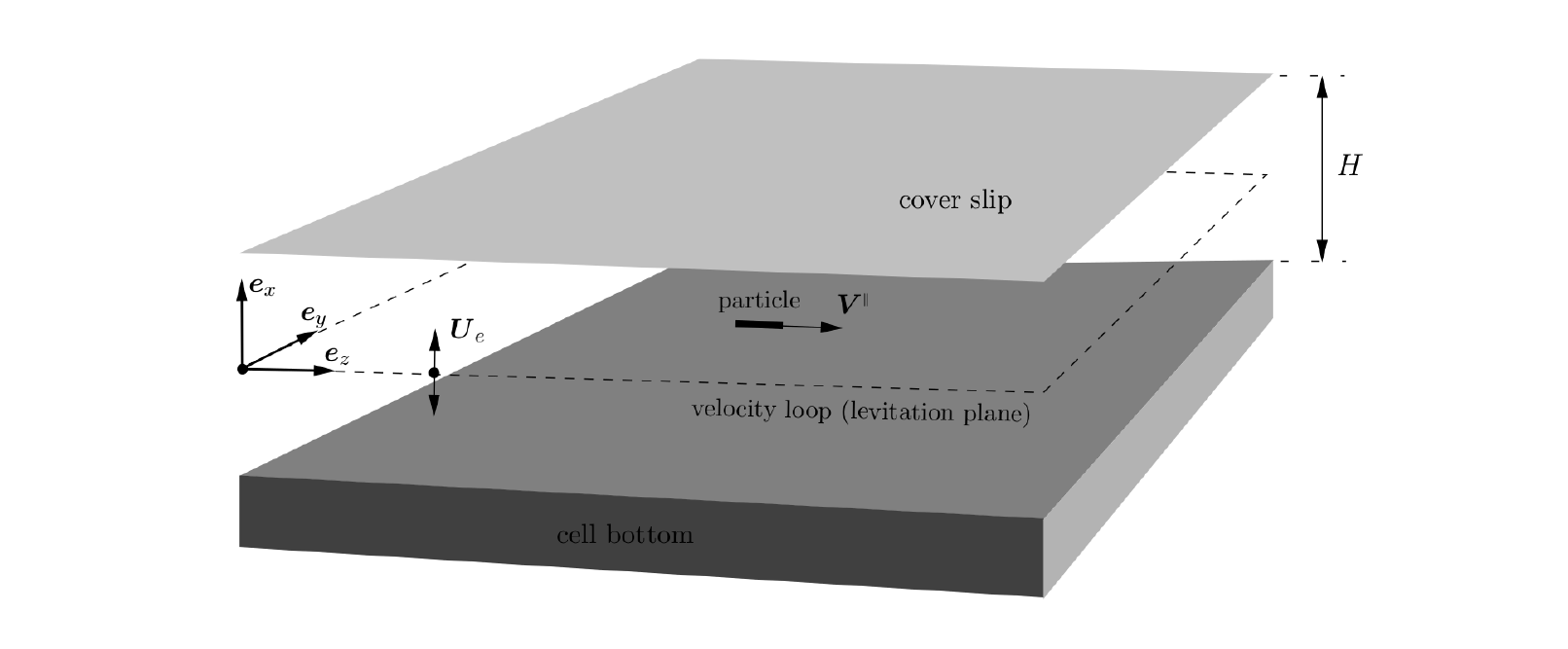}}
\caption{Sketch of the experimental setup in Ref.~\cite{Wang2012}. The thickness of the sample is $H = 180\,\mu$m,
and the first acoustic resonance is found at $\omega/2\pi = 3.7$ MHz. Typical size of the cylindrical rods (resp. spheres) is
3 $\mu$m length $\times$ 300 nm diameter (resp. 2 $\mu$m diameter).}
\label{exp_setup}
\end{figure}

Using our mathematical model, we can  give an order of magnitude of the propulsion speed for this mechanism in  the experimental configuration of 
 Ref.~\cite{Wang2012}, whose setup is  recalled in figure \ref{exp_setup}. Micron-size spherical and cylindrical particles are positioned acoustically 
(radiation pressure)  in the center of a water cell of thickness $H=180$\,$\mu$m, corresponding to a first acoustic resonance of $\omega=3.7$ MHz. 
We note that our theory was derived in the  asymptotic
limit  $\lambda^2 \ll 1$. However, in this experiment, taking a  typical size $R_{\sst 0} = 1$ in water leads to $\lambda^2 = 23$, so our results 
should be understood as providing at best an order of magnitude estimate.

We first consider  the case of metallic  (gold) rods. The typical size of the cylindrical rods is 3 $\mu$m length $\times$ 300 nm diameter 
(slender body). Without further information, we take $\eps = 0.1$  and we consider the mode $k = 1$. In Wang's experiments, the power provided 
to the  fluid by the acoustic forcing is estimated to be lower than 1.25 W$\,$cm$^{-2}$. From the value of this upper bound, and considering 
that the power density $\mathcal{P}$ by surface unit in  the cell can be estimate using $\mathcal{P} \sim \rho\,\hat{U}_e^2\omega^3 H$, the 
amplitude of the fluid oscillations,  $\hat{U}_e/\omega$,
can be estimated at about 2.35 nm. To compute the amplitude of oscillations of the particle relative to the fluid, we have to take the 
inertia of the particle into account (equation \ref{rel_transv_speed}). For gold particles
in water the density ratio is $\beta = 5.18 \times 10^{-2}$, such that, for a frequency parameter $\lambda^2 \sim 23$,
we obtain $a= \hat{V}^{\per}/\omega = 2.1$ nm. 
Introducing the values $\eps = 0.1$, $a = 2.1\times10^{-9}$~m, $\omega = 2\pi\times 3.7\times10^{6}$~s$^{-1}$, $R_{\sst 0} = 10^{-6}$~m 
and $\nu = 10^{-6}$~m$^2\,$s$^{-1}$, and the computed value $v^{\sst(1,1)}_{\sst 1} = 0.11$ in equation (\ref{dimensional_prop_speed}) 
we obtain $V^{\pa} \sim 26\,\mu$m$\,$s$^{-1}$. This value is  lower than the upper bound of $\sim200$~$\mu$m s$^{-1}$ measured experimentally 
in Ref.~\cite{Wang2012}, but at least of the correct order of magnitude given the unknowns in the experimental fit. In particular,  we have 
assumed our shape to be roughly spherical whereas cylinders are known to experience lower viscous drag than their equivalent spheres. The 
degree of geometrical asymmetry in the experiment is also unknown. Note that  in Ref.~\cite{Wang2012} it is mentioned  that metallic spheres 
are sometimes able to swim but no measurements of the speeds are reported.

If we now consider  the case of polymeric rods and spheres, the situation is quite different since these particles, due to their low density,  
show a smaller relative velocity. Polystyrene particles have $\beta = 0.94$ and thus  almost follow the forcing flow with little  relative 
motion. Using the same parameters as above for the fit, we now  obtain $a \sim 0.11$ nm.  The propulsion speed is then found to be smaller 
than the one calculated for metallic particles by two orders of magnitude, which might  explain the experimental observation that  polymeric 
particles do not swim. Recall from equation (\ref{rel_transv_speed}) that the  perturbative flow responsible for the steady streaming and 
the propulsion vanishes for particles of density similar to that of the fluid. In order for this acoustic mechanism to be effective, the  
density of the particle  must be far from the density of the surrounding fluid to ensure a large relative motion and therefore an efficient 
propulsion. 

The predictions of our model are thus in qualitative agreement with the experimental observations. To fully  capture the experimentally-relevant 
limit, a calculation should be carried out in the limit  $\lambda^2 \gg 1$. In that case, the expansions would have to be considered differently 
and the two relevant small parameters would then be the amplitude-to-size ratio, $a/R_{\sst 0}$, instead of the Reynolds number \cite{Riley1966}, 
and the shape parameter (as in the case $\lambda^2 \ll 1$). This limit will be addressed in future work.

\section*{Acknowledgements}
The authors thank the Department of Mechanical and Aerospace Engineering at the University of California San Diego  where  this research was initiated. 
This work was funded in part by the Seventh Framework Programme of the European Union through a Marie Curie grant to EL (Grant PCIG13-GA-2013-618323),
in part by the Direction G\'en\'erale de l'Armement (Grant 2012600091 - Project ERE 12C0020).  

\appendix
\section{Choice of the position of the origin \label{origin_position}}

\begin{figure}[b]
\scalebox{0.9}{\includegraphics{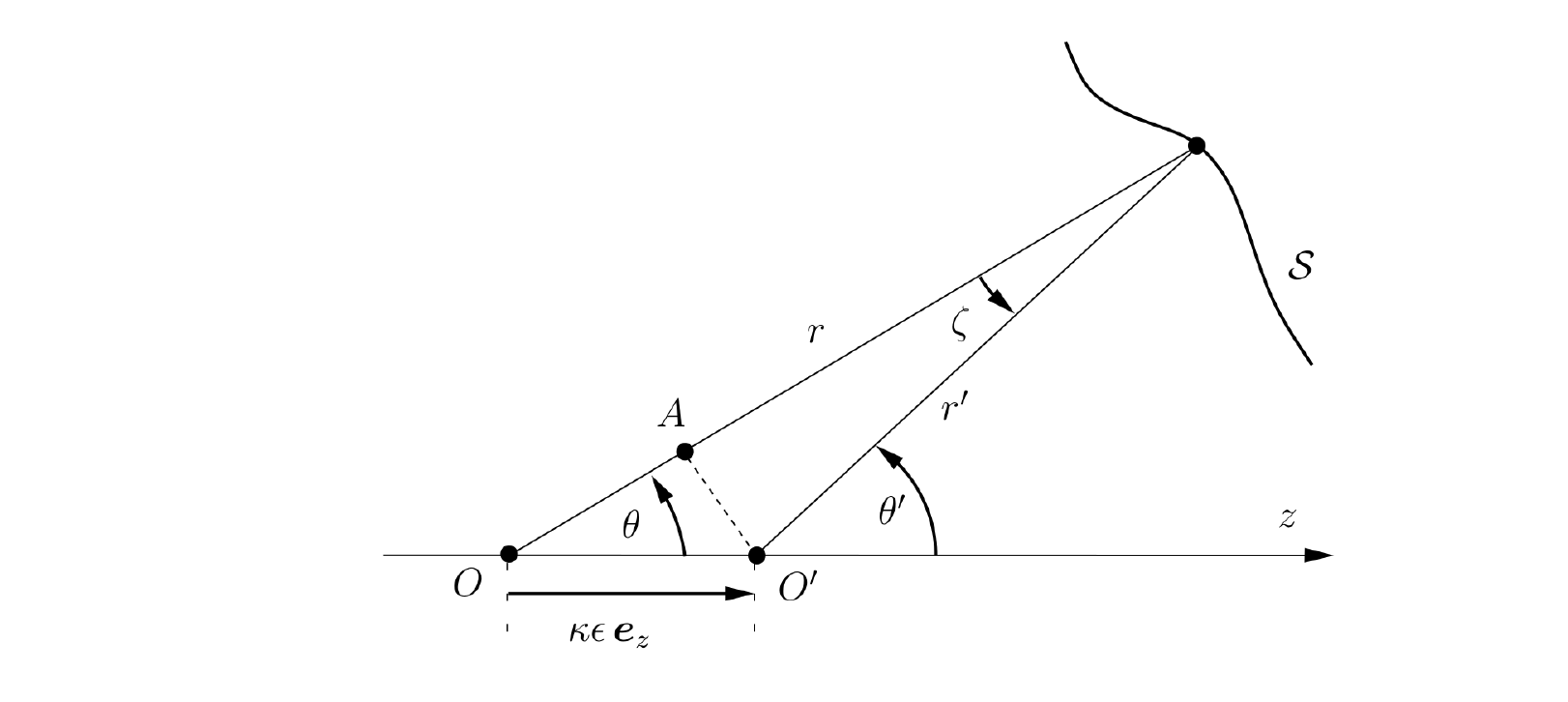}}
\caption{Translation of the origin of the coordinate system along the $z$-axis.}
\label{origin_pos}
\end{figure}

In this Appendix, we investigate how  the polar equation $r = 1 + \eps\,\xi(\theta)$ transforms when the origin of the coordinate 
system is translated along the $z$-axis. The answer will enable us to show that$n = 1$ corresponds to a sphere at order $O(\eps)$ and that, 
consequently, no propulsion can be achieved in this case by symmetry. The propulsion speed would thus not 
depend of the choice made for the position of the origin on the $z$-axis.

We  consider an axisymmetric body of axis $z$, the polar equation of which is given by
$r = 1 + \eps\,\xi(\theta)$. The origin of the coordinates system is then translated
of a quantity $\kappa\eps\,\vecz$ with $\kappa = O(1)$. The new polar coordinates are referred to as
$r'$ and $\theta'$ (see notation in figure \ref{origin_pos}). Knowing that $r\cos\theta - r'\cos\theta'= \kappa\eps$ and $r\sin\theta = r'\sin\theta'$,
we can derive the equality
\begin{equation}
r = r'\left(1+\frac{\kappa\eps}{r'}\cos\theta'\right) + O(\eps^2).
\label{corr_r}
\end{equation}
On the other hand, knowing that $\theta = \theta'-\zeta$ and $\zeta r' = -\kappa\eps\,\sin\theta' + O(\eps^2)$,
we can derive a second equation linking $\theta$ and $\theta'$
\begin{equation}
\theta = \theta' - \frac{\kappa\eps}{r'}\sin\theta' + O(\eps^2).
\label{corr_theta}
\end{equation}
Introducing equations (\ref{corr_r})
and (\ref{corr_theta}) into equation (\ref{polar_adim}) leads to
\begin{equation}
r' = 1 + \eps\,[\xi(\theta')-\kappa\cos\theta'] + O(\eps^2).
\label{polar_new}
\end{equation}
If we consider the case $n = 1\;(k=0)$
($\xi(\theta) = \cos\theta$), and choose to translate the origin of a quantity $\eps\vecz$ ($\kappa = 1$),
the polar equation of the surface in terms of the new polar coordinates reduces to
\begin{equation}
r' = 1 + O(\eps^2).
\end{equation}
In other words, the case $n=1$ is nothing but a simple translation of vector $\eps\vecz$ of the equivalent-volume
sphere (the polar equation in terms of new coordinates derived up to  $O(\eps^2)$ shows that the case $n=1$ actually corresponds to an oblate spheroid). 

Let us now consider an arbitrary axisymmetric shape written as
\begin{equation}
\xi(\theta) = \sum_{n=1}^{\infty} \zeta_n \cos(n\theta).
\end{equation}
Introducing this Fourier series in equation (\ref{polar_new}), one gets
\begin{equation}
r' = 1 + \eps \left[\sum_{n=1}^{\infty} \zeta_n \cos(n\theta') - \kappa\cos\theta'\right] + O(\eps^2),
\end{equation}
and only the amplitude of the term $n=1$ is affected by the change of origin. Since the total propulsion speed is linear with respect
to the shape function and we saw that the term $n=1$, corresponding to a sphere,  has no contribution in the propulsion speed, we conclude 
that the total propulsion speed does not depend on the position of the origin along the $z$ axis. Note that new shape function involved in 
the polar equation (\ref{polar_new}) in terms of new coordinates satisfies also the integral condition from equation (\ref{cond_1}) provided 
that the initial shape function does satisfy this condition.  
 
\section{Oscillations of a sphere in a viscous fluid \label{oscillating_sphere}}

We recall in this Appendix the expression of the flow produced by a sphere of radius $R_{\sst 0}$
oscillating with a velocity $\vecV^{\per} = \hat{\vecV}^{\per} \expo^{-\icomps \omega T} $ in a viscous fluid. The amplitude
of the oscillations, $a = |\hat{\vecV}^{\per}|/\omega$, is assumed to be smaller
than the radius $R_{\sst 0}$ by at least one order of magnitude, $a/R_{\sst 0} \ll 1$.
The problem is governed by the Navier-Stokes (NS) equations
\begin{gather}
\rho\left[\frac{\partial \vecU^{\sst 0}}{\partial t} + (\vecU^{\sst 0}\cdot\Nab)\vecU^{\sst 0}\right] =
- \Nab P^{\sst 0} +\mu\,\Nab^2\vecU^{\sst 0},\label{NS11s}\\
\Nab\cdot\vecU^{\sst 0} = 0,\label{NS21s}
\end{gather}
where $\vecU^{\sst 0}$ and $P^{\sst 0}$ are the dimensional velocity and pressure fields.
Before completing the previous system by a suitable set of boundary conditions, let us
make the NS equations dimensionless. For sake of simplicity, we adopt a particular choice of non dimensionalization
for distances. Instead of the radius of the colloid, we choose the quantity $\alpha = (\icomp\nu/\omega)^{1/2}$, which
is the distance over which the viscosity diffuses. Furthermore, we choose $\omega^{-1}$,
$\hat{V}^{\per} = a\,\omega$, $\mu \alpha \hat{V}^{\per}$ as typical time, velocity, and pressure.
Then, the dimensionless NS equations simplify into the linear unsteady Stokes equations
\begin{gather}
\hat{\vecu}^{\sst 0} = -\Nabt \hat{p}^{\sst 0} + \Nabt^2\hat{\vecu}^{\sst 0},\label{NS12s}\\
\Nabt \cdot \hat{\vecu}^{\sst 0}  = 0,\label{NS22s}
\end{gather}
where $\hat{\vecu}^{\sst 0}$ are $\hat{p}^{\sst 0}$ are the dimensionless Fourier components (of dimensionless frequency 1)
of the velocity and pressure fields and $\Nabt$ is the new gradient operator. In equation (\ref{NS12s}),
the nonlinear term has been neglected since it is smaller than any other by a factor $a/R_{\sst 0}$. In the following, the position vector 
is referred to as $\vecraux$, its norm being denoted by $\raux$.
The boundary conditions in the frame of reference of the laboratory are given by 
\begin{align}
\hat{\vecu}^{\sst 0} & = \hat{\vecv}^{\per}\;\;\;\mbox{on the surface of the sphere},\label{osc_sph_BC1_bis}\\
\hat{\vecu}^{\sst 0} & \rightarrow 0 \;\;\mbox{for}\;\;|\vecraux| \rightarrow \infty,\label{osc_sph_BC2_bis}
\end{align}
where $\hat{\vecv}^{\per}$ is the unit vector aligned with the direction of oscillation. Following classical work  \cite{Kim&Karrila}, 
we can write the solution in the form
\begin{equation}\label{uG}
\hat{\vecu}^{\sst 0} = \frac{3}{4}\lambda_{\sst 0}\,\hat{\vecv}^{\per}\cdot[\Lambda_{\sst 0} +
\Lambda_{\sst 1} \lambda_{\sst 0}^2 \Nabt^2]\vecG(\vecraux),
\end{equation}
where $\lambda_{\sst 0} = \expo^{-\icomps \pi/4} \lambda = \expo^{-\icomps \pi/4} (R_{\sst 0}^2\omega/\nu)^{1/2}$, and where
$\Lambda_{\sst 0}$ and $\Lambda_{\sst 1}$ are quantities to be determined using 
the boundary conditions. In  equation \eqref{uG}, the fundamental solution of the unsteady
Stokes equation $\vecG$ has been used. It is given by
\begin{equation}
\vecG(\vecraux) = g(\raux)\,\vecraux \vecraux + h(\raux)\,\vecdelta,
\end{equation}
where
\begin{gather}
g(\raux) = \frac{2}{\raux^5}[3 - (3 + 3 \raux + \raux^2)\expo^{-\raux}],\\
h(\raux) = \frac{2}{\raux^3}[(1+\raux+\raux^2)\expo^{-\raux}-1],
\end{gather}
and $\vecdelta$ is the unit tensor.
Using the two identities
\begin{gather}
\frac{\partial^2}{\partial \xaux_k^2} [f(\raux) \xaux_i \xaux_j] = 
\xaux_i\xaux_j\,(\tilde{D}_{\sst 2}+ 4 \tilde{D}_{\sst 1}) f(\raux)+ 2 f(\raux)\,\delta_{ij},\\
\frac{\partial^2}{\partial \xaux_k^2} [f(\raux) \delta_{ij}] = \delta_{ij}\,\tilde{D}_{\sst 2} f(\raux),
\end{gather}
where
\begin{equation}
\tilde{D}_{\sst 1} = \frac{1}{\raux}\frac{\partial}{\partial \raux}
\;\;\mbox{and}\;\;
\tilde{D}_{\sst 2} = \frac{1}{\raux^2}\frac{\partial}{\partial \raux}\left(\raux^2\frac{\partial}{\partial \raux}\right),
\end{equation}
we can easily derive the unsteady velocity flow produced by the oscillation of a sphere in a viscous
fluid as
\begin{equation}
\hat{\vecu}^{\sst 0}(\vecraux) = \frac{3}{2} \hat{\vecv}^{\per} \cdot \frac{\lambda_{\sst 0}}{\raux^3}\bigg[
g_{\sst 0}(\raux)\frac{\vecraux \vecraux}{\raux^2} + h_{\sst 0}(\raux)\vecdelta\bigg].
\label{flow_u0}
\end{equation}
where
\begin{align}
& g_{\sst 0}(\raux) = 3\,\Lambda_{\sst 0} - (\Lambda_{\sst 0}+\lambda_{\sst 0}^2 \Lambda_{\sst 1}) 
(3+3\raux+\raux^2)\expo^{-\raux},\\
& h_{\sst 0}(\raux)= -\Lambda_{\sst 0} +(\Lambda_{\sst 0}+\lambda_{\sst 0}^2\Lambda_{\sst 1})(1+\raux+\raux^2)\expo^{-\raux}.
\end{align}
The boundary condition in equation (\ref{osc_sph_BC1_bis}) at the surface of the sphere provides the explicit form
of $\Lambda_{\sst 0}$ and $\Lambda_{\sst 1}$ and we obtain
\begin{equation}
\Lambda_{\sst 0} = 1 + \lambda_{\sst 0} + \frac{\lambda_{\sst 0}^2}{3}\;\;
\mbox{and}\;\;\Lambda_{\sst 1} = \lambda_{\sst 0}^{-2}(\expo^{\lambda_{\sst 0}}-\Lambda_{\sst 0}).\label{Lambda0Lambda1}
\end{equation}

The solution in equation (\ref{flow_u0}) allows to derive the partial
derivative of the velocity with respect to the radial distance at the particle surface. Using the dimensionless variable $r = 
\lambda_{\sst 0}^{-1}\raux$, i.e.~the dimensionless variable obtained by choosing $R_{\sst 0}$ as typical length, we obtain
\begin{equation}
\left.\frac{\partial \hat{\vecu}^{\sst 0}}{\partial r}\right|_{r=1} = 
-\frac{3}{2}(1+\lambda_{\sst 0})\hat{\vecv}^{\per}\cdot(\vecdelta-\vecn\vecn),\label{r_derivative_1}
\end{equation}
where $\vecn$ is the outwards unit vector normal to the surface of the spherical particle.

\section{Transverse oscillations of a near-sphere in a viscous fluid \label{transv_oscillation}}

In this Appendix, we first detail the method used to write the boundary conditions on the surface gradient and surface curl in terms 
of associated Legendre functions. We then obtain explicit expressions for   the components $\hat{u}_r^\eps$, $\hat{u}_\theta^\eps$, 
$\hat{u}_\phi^\eps$ of the velocity field
$\hat{\vecu}^\eps$.

\subsection{Boundary conditions in terms of associated Legendre functions}

Introducing the polar and azimutal components of $\hat{\vecu}^\eps$ at the surface given by equation (\ref{stokes_BC1}) into the
right-hand side  of equations (\ref{surf_div}) and (\ref{surf_curl}), we obtain
\begin{gather}
- \Nab_s \cdot \hat{\vecu}^\eps = K \cos\phi\,[n\sin(n\theta)\cos\theta + 2 \cos(n\theta) \sin\theta],\\
\vecer\cdot\Nab_s\times \hat{\vecu}^\eps = K \sin\phi\,n \sin(n\theta).
\end{gather}
Recalling that $n = 2k+1$ $(k \ge 1)$, one can show that the two previous equations can be put in the form
\begin{gather}
- \Nab_s \cdot \hat{\vecu}^\eps = K \cos\phi\,\sin\theta\,\sum_{q=0}^{k}A_{\sst 2q+1}\cos^{2q+1}\theta,\\
\vecer\cdot\Nab_s\times\hat{\vecu}^\eps = K \sin\phi\,\sin\theta\,\sum_{q=0}^{k}A_{\sst 2q}\cos^{2q}\theta,
\end{gather}
with
\begin{gather}
A_{\sst 2q+1} = (-1)^{k-q}\sum_{m=0}^{q}[(2k+1)C^{\sst 2k+1}_{\sst 2m}+2C^{\sst 2k+1}_{\sst 2m+1}]C^{\sst k-m}_{\sst q-m},\label{A_2q+1}\\
A_{\sst 2q} = (-1)^{k-q}(2k+1)\sum_{m=0}^{q} C^{\sst 2k+1}_{\sst 2m}C^{\sst k-m}_{\sst q-m}.\label{A_2q}
\end{gather}
Now, after noticing that associated Legendre functions of order 1, and of even and odd degree,
are of the form
\begin{gather}
P^{\sst 1}_{\sst 2(q+1)}(\cos\theta)  = \sin\theta\,\sum_{l=0}^{q}a_{\sst 2l+1}^{\sst 2(q+1)}\cos^{2l+1}\theta,\\
\mbox{and}\;\;P^{\sst 1}_{\sst 2q+1}(\cos\theta)  = \sin\theta\,\sum_{l=0}^{q}a_{\sst 2l}^{\sst 2q+1}\cos^{2l}\theta,
\end{gather}
one can rewrite $- \Nab_s \cdot \hat{\vecu}^\eps$ and $\vecer\cdot\Nab_s\times\hat{\vecu}^\eps$ in the form
of a sum of associated Legendre functions 
\begin{gather}
- \Nab_s \cdot \hat{\vecu}^\eps = K \cos\phi\,\sum_{q=0}^{k} B_{\sst 2(q+1)} P^{\sst 1}_{\sst 2(q+1)}(\cos\theta) ,\\
\vecer\cdot\Nab_s\times \hat{\vecu}^\eps = K \sin\phi\,\sum_{q=0}^{k} B_{\sst 2q+1} P^{\sst 1}_{\sst 2q+1}(\cos\theta),
\end{gather}
where the coefficients $B_{\sst 2(q+1)}$ and $B_{\sst 2q+1}$ are the respective solutions of the two  systems
\begin{gather}
\sum_{q=l}^{k} B_{\sst 2(q+1)} a^{\sst 2(q+1)}_{\sst 2l+1} = A_{\sst  2l+1}\;\;(l = 0,\cdots,k),\\
\sum_{q=l}^{k} B_{\sst 2q+1} a^{\sst 2q+1}_{\sst 2l} = A_{\sst 2l}\;\;(l = 0,\cdots,k).
\end{gather}
This derivation allows to use equations (\ref{surf_div_BC}), (\ref{surf_curl_BC}) as the suitable forms of the boundary
condition at $r=1$.

\subsection{Expressions of the components of the corrective velocity field $\hat{\vecu}^{\eps}$}
Some algebra leads to
\begin{eqnarray}
\hat{u}^\eps_r &=& K\cos\phi\,\sum_{q=0}^{k}\frac{U_{\sst 2(q+1)}(r)}{r}\,P^{\sst 1}_{\sst 2(q+1)}(\cos\theta),\\
\hat{u}^\eps_\theta &=& 
\frac{K}{2}\,r \cos\phi\,\sum_{q=0}^{k}\frac{1}{q+1}\bigg[
\frac{1}{2q+3} V_{\sst 2(q+1)}(r)\,\frac{d P^{\sst 1}_{\sst 2(q+1)}(\cos\theta)}{d \theta}\hspace{2cm}\nonumber\\
&&\hspace{3.5cm} + \frac{1}{2q+1} Y_{\sst 2q+1}(r)\,\frac{P^{\sst 1}_{\sst 2q+1}(\cos\theta)}{\sin\theta}\bigg],\\
\hat{u}^\eps_\phi &=& 
-\frac{K}{2}\,r \sin\phi\,\sum_{q=0}^{k}\frac{1}{q+1}\bigg[
\frac{1}{2q+3} V_{\sst 2(q+1)}(r)\,\frac{P^{\sst 1}_{\sst 2(q+1)}(\cos\theta)}{\sin\theta}\hspace{2cm}\nonumber\\
&&\hspace{3.5cm} + \frac{1}{2q+1} Y_{\sst 2q+1}(r)\,\frac{d P^{\sst 1}_{\sst 2q+1}(\cos\theta)}{d \theta}\bigg],
\end{eqnarray}
where
\begin{gather}
V_{\sst 2(q+1)}(r) = \frac{d}{dr}\left(\frac{U_{\sst 2(q+1)}}{r}\right) + \frac{2 U_{\sst 2(q+1)}}{r^2},\\
Y_{\sst 2q+1}(r) = \frac{X_{\sst 2q+1}}{r},
\end{gather}
and where the quantities $d P^{\sst 1}_{\sst l}(\cos\theta)/d\theta$ and $P^{\sst 1}_{\sst l}(\cos\theta)/\sin\theta$ can
be calculated using the identities
\begin{gather}
\frac{d P^{\sst 1}_{\sst l}(\cos\theta)}{d \theta} = 
-l(l+1) P^{\sst 0}_{\sst l} + \frac{1}{2}\cos\theta[P^{\sst 2}_{\sst l+1} +l(l+1) P^{\sst 0}_{\sst l+1}],\\
\frac{P^{\sst 1}_{\sst l}(\cos\theta)}{\sin \theta} = - \frac{1}{2}\cos\theta[P^{\sst 2}_{\sst l+1} +l(l+1) P^{\sst 0}_{\sst l+1}].
\end{gather}

\section{Steady translation of an axisymmetric near-sphere in a viscous fluid \label{axial_translation}}

The solution to the problem of an axisymmetric near-sphere translating in a purely viscous fluid at constant speed $\bar{\vecv}$ along its
axis of symmetry (here the $z$-axis) is  known. We remind here some 
useful results in the case of a shape function
of the form $\xi(\theta) = \cos n\theta$ with $n=2k+1$. The pressure-velocity field $(\bar{\vecu},\,\bar{p})$ satisfies the
dimensionless Stokes equations
\begin{gather}
-\Nab \bar{p} + \Nab^2\bar{\vecu} = 0,\label{Stokes_eq_1}\\
\Nab \cdot \bar{\vecu}  = 0.\label{Stokes_eq_2}
\end{gather}
Writing the fields $\bar{\vecu}$ and $\bar{p}$ in the form
\begin{align}
\bar{\vecu} = \bar{\vecu}^{\sst 0} + \eps\,\bar{\vecu}^\eps + O(\eps^2) \label{expand_uaux_Re0_eps}\\
\bar{p} = \bar{p}^{\sst 0} + \eps\,\bar{p}^\eps + O(\eps^2) \label{expand_paux_Re0_eps},
\end{align}
where $\bar{\vecu}^{\sst 0}$ and $\bar{p}^{\sst 0}$ are the velocity and presssure fields
induced by the steady translation of a sphere in a purely viscous fluid, and where
$\bar{\vecu}^{\eps}$ and $\bar{p}^{\eps}$ are the corrective fields due to the non sphericity of the
particle. The fields $\bar{\vecu}^{\sst 0}$ and $\bar{p}^{\sst 0}$ are the classical Stokes solution for flow past a sphere.
The fields $\bar{\vecu}^{\eps}$ and $\bar{p}^{\eps}$ also satisfy the Stokes equations
\begin{gather}
-\Nab \bar{p}^\eps + \Nab^2\bar{\vecu}^\eps = 0,\label{Stokes_eq_aux_1}\\
\Nab \cdot \bar{\vecu}^\eps  = 0,\label{Stokes_eq_aux_2}
\end{gather}
and $\bar{\vecu}^{\eps}$ must vanish at infinity, that is to say
\begin{equation}
\bar{\vecu}^\eps \rightarrow \vecO\;\;\mbox{for}\;|\vecr|\rightarrow\infty.\label{BC_aux_inf}
\end{equation}
As in the case of a the transverse oscillations of a near-sphere in a
viscous fluid, the boundary condition at the particle surface takes the simple Taylor-expansion form
\begin{equation}
\bar{\vecu}^\eps|_{r=1} = -\xi(\theta)\left.\frac{\partial \bar{\vecu}^{\sst 0}}{\partial r}\right|_{r=1}. \label{aux_BC}
\end{equation}
The derivative of $\bar{\vecu}^{\sst 0}$ at $r=1$ has a form similar to the steady limit of equation (\ref{r_derivative_1})
\begin{equation}
\left.\frac{\partial \bar{\vecu}^{0}}{\partial r}\right|_{r=1} = 
-\frac{3}{2}\,\bar{\vecv}\cdot(\vecdelta-\vecn\vecn).\label{r_derivative_2}
\end{equation}
Introducing explicitly the direction of the translation speed $\bar{v} = \vecz$, the boundary condition, equation (\ref{aux_BC}), 
becomes
\begin{equation}
\bar{\vecu}^\eps|_{r=1} = \bar{K}\,\xi(\theta)\left(
\begin{array}{c}
0\\
-\sin\theta\\
0
\end{array}
\right),
\label{Stokes_BC_aux}
\end{equation}
where $\bar{K} = 3/2$.

The method then used to derive   the solution to equations (\ref{Stokes_eq_aux_1}-\ref{Stokes_eq_aux_2}) is not different
from the method used in section \ref{zeroth_order} and Appendix \ref{transv_oscillation} to derive the
corrective field $\bar{\vecu}^\eps$. We keep  the continuity condition on the radial component of the velocity as
\begin{equation}
\bar{u}_r^\eps = 0 \;\;\mbox{at}\;r=1 \label{rad_vel_BC_aux},
\end{equation}
and use continuity conditions on the surface divergence and the surface curl at the particle surface instead of continuity conditions 
on the polar and azimuthal component of the velocity. Introducing equation 
(\ref{Stokes_BC_aux}) into equations (\ref{surf_div}) and (\ref{surf_curl}) lead to the new boundary conditions at $r = 1$ 
\begin{gather}
- \Nab_s \cdot \bar{\vecu}^\eps = \bar{K} \cos\phi\,[- n\sin(n\theta)\sin\theta + 2 \cos(n\theta) \cos\theta],\label{surf_div_aux}\\
\vecer\cdot\Nab_s\times\bar{\vecu}^\eps = 0.\label{surf_curl_aux}
\end{gather}
Recalling that $n = 2k+1$ $(k \ge 1)$, one can show that the first of the previous equations can be written as
\begin{gather}
- \Nab_s \cdot \bar{\vecu}^\eps = \bar{K} \sum_{q=0}^{k+1}\bar{A}_{\sst 2q}\cos^{2q}\theta,
\label{surf_div_aux_1}
\end{gather}
with
\begin{gather}
\bar{A}_{\sst 0} = (2k+1)\,(-1)^{k+1},\label{A_0_aux}\\
\bar{A}_{\sst 2q} = (-1)^{k-q+1}\sum_{m=0}^{q}\{[(2C^{\sst 2k+1}_{\sst 2m+1} + (2k+1)C^{\sst 2k+1}_{\sst 2m}]C^{\sst k-m}_{\sst q-m}
\hspace{3cm}\nonumber\\
\hspace{1cm} + (2k+1)C^{\sst 2k+1}_{\sst 2m}C^{\sst k-m}_{\sst q-m-1}\} + (-1)^{k-q+1}(2k+1)C^{\sst 2k+1}_{\sst 2q}\nonumber\\[2mm]
\hspace{8cm}\mbox{for}\;\;q = 0\cdots k,\label{A_2(q+1)_aux}\\
\bar{A}_{\sst 2(k+1)} = \sum_{m=0}^{k} [2C^{\sst 2k+1}_{\sst 2m+1} + (2k+1)C^{\sst 2k+1}_{\sst 2m}].\label{A_2(p+1)_aux}
\end{gather}
The associated Legendre functions of order 0, and of even degree,
are of the form
\begin{equation}
P^{\sst 0}_{\sst 2q}(\cos\theta)  = \sum_{l=0}^{q}\bar{a}_{\sst 2l}^{\sst 2q}\cos^{2l}\theta,
\end{equation}
so that one can rewrite $- \Nab_s \cdot \bar{\vecu}^\eps$ as a sum of associated Legendre functions
\begin{equation}
- \Nab_s \cdot \bar{\vecu}^\eps = \bar{K}\,\sum_{q = 1}^{k+1} \bar{B}_{\sst 2q}P^{\sst 0}_{\sst 2q}(\cos\theta),
\label{surf_div_aux_2}
\end{equation}
where the constants $\bar{B}_{\sst 2q}$ are solutions of the system
\begin{equation}
\sum_{q=l}^{k+1}\bar{B}_{\sst 2q}\bar{a}^{\sst 2q}_{\sst 2l} = \bar{A}_{\sst 2l}\;\;\;(l = 1 \cdots k+1).
\end{equation}
Note that in equation (\ref{surf_div_aux_2}), $P^{\sst 0}_{\sst 0}$ has no contribution ($\bar{B}_{\sst 0} = 0$), since,
using equation (\ref{surf_div_aux}), one can demonstrate that
\begin{equation}
\int_{0}^{\pi} \Nab_s \cdot \bar{\vecu}^\eps \sin\theta\,d\theta = 0. 
\end{equation}

The general form of the solution to  equations (\ref{Stokes_eq_1}-\ref{Stokes_eq_2}) has been given in Refs.~\cite{Lamb,Kim&Karrila}. 
With the  boundary conditions in equations  (\ref{BC_aux_inf})
and (\ref{surf_curl_aux}) taken into account, that solution reduces to
\begin{equation}
\bar{\vecu}^\eps = \sum_{q=1}^{k+1}\bigg[\frac{(1-q)\,r^2}{2q (4q-1)}\Nab\varphi_{\sst 2q} 
+ \frac{(2q+1)\,\vecr}{2q (4q-1)}\varphi_{\sst 2q}\bigg] + \sum_{q=1}^{k+1} \Nab\psi_{\sst 2q},\label{Stokes_gene}
\end{equation}
where
\begin{align}
& \varphi_{\sst 2q} = \bar{K}\,\bar{\alpha}_{\sst 2q}\,r^{-(2q+1)}\,P^{\sst 0}_{\sst 2q}(\cos\theta),\label{phi_2q}\\
& \psi_{\sst 2q} = \bar{K}\,\bar{\beta}_{\sst 2q}\,r^{-(2q+1)}\,P^{\sst 0}_{\sst 2q}(\cos\theta).\label{psi_2q}
\end{align}

Introducing equations (\ref{phi_2q}) and (\ref{psi_2q}) into equation (\ref{Stokes_gene}), and recalling that $-\Nab_s \cdot \bar{\vecu}^\eps =
\partial_r\bar{u}^\eps$, we can derive explicit expressions of the the radial component of the velocity and surface divergence at $r=1$ as
\begin{gather}
\bar{u}^\eps = \bar{K}\,\sum_{q=1}^{k+1}\left[\frac{2q+1}{2(4q-1)}\bar{\alpha}_{\sst 2q} - (2q+1)\bar{\beta}_{\sst 2q}\right]
\,P^{\sst 0}_{\sst 2q}(\cos\theta),\label{surf_div_aux_3}\\
-\Nab_s \cdot \bar{\vecu}^\eps = \bar{K}\,\sum_{q=1}^{k+1}\left[\frac{q(2q+1)}{2 (1-4q)}\bar{\alpha}_{\sst 2q} + 
2(2q+1)(q+1)\bar{\beta}_{\sst 2q}\right]\,P^{\sst 0}_{\sst 2q}(\cos\theta).
\end{gather}
Using equations (\ref{rad_vel_BC_aux}) and  (\ref{surf_div_aux_2}) we then obtain the system
\begin{gather}
\frac{1}{2(4q-1)}\bar{\alpha}_{\sst 2q} - \bar{\beta}_{\sst 2q} = 0,\\
(2q+1)\left[\frac{q}{1-4q}\bar{\alpha}_{\sst 2q} + 2(q+1)\bar{\beta}_{\sst 2q}\right] = \bar{B}_{\sst 2q},
\end{gather}
which gives $\bar{\alpha}_{\sst 2q}$ and $\bar{\beta}_{\sst 2q}$ explicitly as
\begin{gather}
\bar{\alpha}_{\sst 2q} = \frac{4q-1}{2q+1}\,\bar{B}_{\sst 2q},\\
\bar{\beta}_{\sst 2q}= \frac{1}{2(2q+1)}\,\bar{B}_{\sst 2q}.
\end{gather}
All three  components of the  velocity field are finally given by
\begin{eqnarray}
\bar{u}_r^\eps&= & \bar{K}\,\sum_{q=1}^{k+1}(2q+1)r^{-2(q+1)}\left[\frac{\alpha_{\sst 2q}}{2(4q-1)}r^2
- \beta_{\sst 2q}\right]\,P^{\sst 0}_{\sst 2q}(\cos\theta),\\
\bar{u}_\theta^\eps &=& \bar{K}\,\sum_{q=1}^{k+1} r^{-2(q+1)}\left[\frac{(1-q)\alpha_{\sst 2q}}{2q(4q-1)}r^2 + \beta_{\sst 2q}\right]
P^{\sst 1}_{\sst 2q}(\cos\theta),\\[2mm]
\bar{u}_\phi &=& 0.
\end{eqnarray}

\section{Propelling flows : dipolar and quadripolar contributions \label{dipolar_quadrupolar}}
\label{lastApp}
In the present work, the transverse propulsion speed (normal to the flow direction) has been calculated in a setup similar to that of  Riley's \cite{Riley1966} for a particle immersed in a uniformly oscillating flow. Because our particles are asymmetric, the inertially-rectified stresses do not in general average to zero, and a net force can be induced normal to the flow direction.  If, however,  the particle is not located at a pressure node (velocity loop) of a standing wave of wave vector $\vecK_{\sst 0} = K_{\sst 0}\,\vecx$, the assumption of a uniform
forcing velocity field no longer holds. The surrounding velocity field would then contain  a linear component, leading to additional dipolar streaming flows  which would have to be properly quantified. The objective of the present Appendix is (i) to detail the respective origins of the dipolar and quadrupolar streaming flows and  (ii) to explain why the dipolar contribution to the global streaming vanishes when the average position of the particle gets closer to the pressure node.

Consider a spherical particle (the exact shape  is not relevant) at a position $X_{\sst 0}$,  in a plane standing wave of the form
$\vecU_e(X) = U_e\,\vecx = \hat{U}_e\sin(K_{\sst 0} X)\,\expo^{-\icomps \omega T}\,\vecx$, where the wavenumber $K_{\sst 0}$ of the wave is the ratio between the pulsation, $\omega$, and the speed of sound, $c$. For the sake of simplicity, we drop the factor $\expo^{-\icomps \omega T}$ in the following.  The incident velocity field can be expanded in the vicinity of the average position $X_{\sst 0}$ of the particle, which yields
\begin{equation}
U_e (X) = \hat{U}_e\,[\sin(K_{\sst 0} X_{\sst0})- 
K_{\sst 0} (X-X_{\sst0}) \cos(K_{\sst 0} X_{\sst0})] + O(K_{\sst 0}R_{\sst 0}^2).
\label{soundwave_1}
\end{equation}
Using equation (\ref{rel_transv_speed}) for the particule velocity (in the laboratory frame), the velocity
field seen by the particle in its own frame of reference becomes
\begin{equation}
U_e(X) = \varpi \hat{U}_e \sin(K_{\sst 0} X_{\sst 0}) - \hat{U}_e K_{\sst 0}(X-X_{\sst0})\,\cos(K_{\sst 0} X_{\sst0}) + O(K_{\sst 0} R_{\sst 0}^2),
\label{velocity_particle_rf}
\end{equation}
where
\begin{equation}
\varpi = \frac{2}{9}\frac{\beta - 1}{\beta}\lambda^2,
\label{Psi}
\end{equation}
and $\lambda \ll 1$. Note that in order to derive (\ref{velocity_particle_rf}) and (\ref{Psi}), we made the assumption that the particle displacement was small compared to $R_{\sst 0}$. 
 
By taking $R_{\sst 0}$, $\varpi \hat{U}_e$ as typical distance and velocity, the field $U_e$ can be written in the following dimensionless form
\begin{equation}
u_e(x) = \icomp\,\sin\,k_{\sst 0} x_{\sst0} - k_\varpi \,(x-x_{\sst0})\,\cos\, k_{\sst 0} x_{\sst0} + O(k_{\sst 0}^2),
\label{expand_lin_incident_flow}
\end{equation}
\begin{figure}
\begin{center}
\scalebox{0.9}{\includegraphics{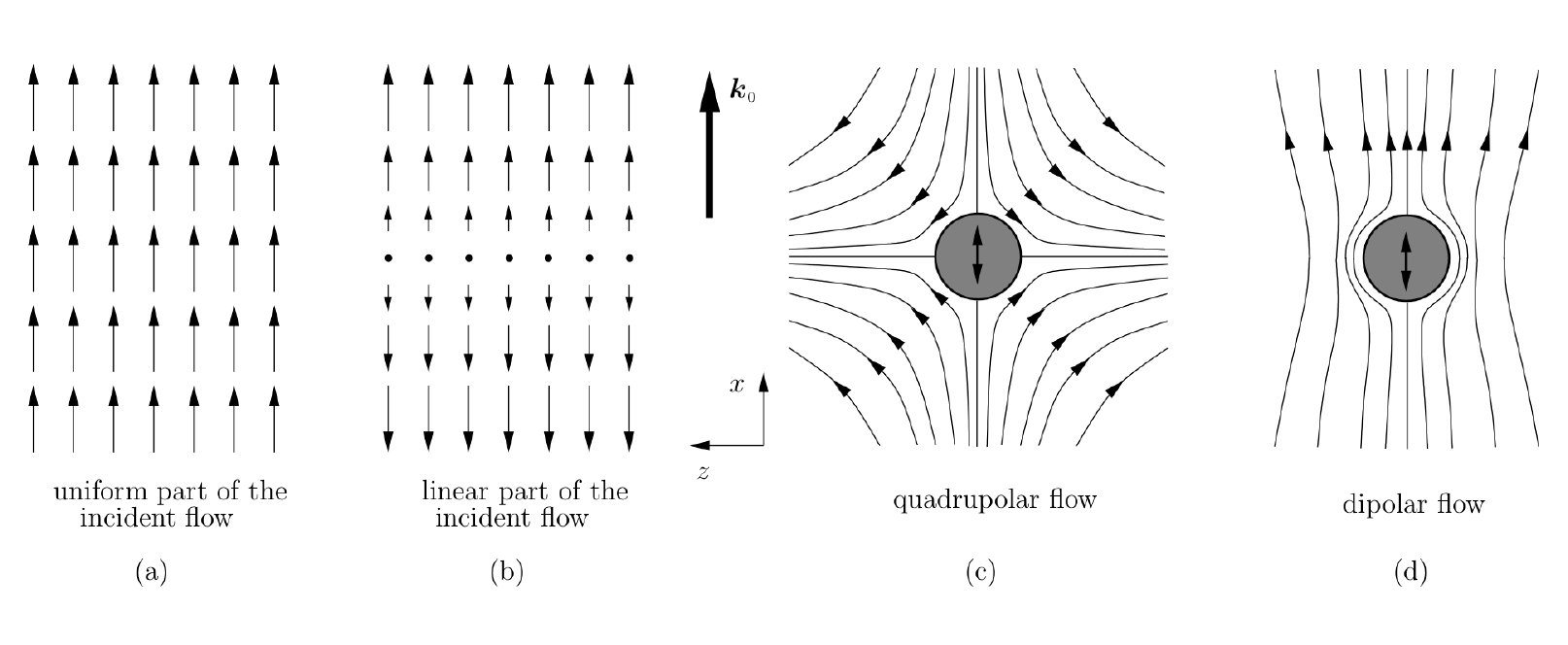}}
\end{center}
\caption{(a) Uniform antisymmetric part of the flow; (b) Linear
symmetric part of the incident flow; (c) Quadrupolar flow in the classic situation of a particle oscillating in a fluid at rest, 
considered by Riley \cite{Riley1966}; (d) dipolar flow in the case of a particle displaced from the pressure node \cite{Danilov2000,Gorkov1962}.}
\label{dip_quad}
\end{figure}
where $k_\varpi = K_{\sst 0} R_{\sst 0}/\varpi$ and $k_{\sst 0} = K_{\sst 0} R_{\sst 0}$. 
In the reference frame of the particle, the incident dimensionless field is therefore the sum of a order-one uniform field of
amplitude $\sin\,k_{\sst 0} x_{\sst0}$, and a linear component of amplitude $k_\varpi \cos\, k_{\sst 0} x_{\sst0}$
(see figure \ref{dip_quad}a and \ref{dip_quad}b).

Let us now consider the Navier-Stokes equations in a dimensionless form. After choosing the quantities $R_{\sst 0}$, $\varpi\,\hat{U}_e$,
$\omega^{-1}$, $\mu \varpi \hat{U}_e/R_{\sst 0}$ and $\rho_{\sst 0}\varpi \hat{U}_e R_{\sst 0} \omega/c^2$
as typical length, velocity, time, stress and density magnitude, and writing the density as the sum of a mean value,  
$\rho_{\sst 0}$ and a deviation, $\rho$, the compressible Navier-Stokes equations takes the form
\begin{gather}
(1 + \varepsilon k_{\sst 0}^2 \rho)\lambda^2\bigg[\frac{\partial \vecu}{\partial t} + 
\varepsilon\,(\vecu\cdot\Nab)\vecu\bigg]  = \Nab\cdot\sigma,\label{NS11comp}\\
k_{\sst 0}^2\frac{\partial \rho}{\partial t} + \Nab\cdot\vecu +  \varepsilon k_{\sst 0}^2 \Nab\cdot(\rho \vecu) = 0,\label{NS21comp}\\
p = \rho,\label{compressibility}
\end{gather}
where $\vecu$, $p$ and $\vecs$ are the velocity, pressure and stress fields and $\veps = \varpi \hat{U}_e/(\omega R_{\sst 0})$.
In the small-$\varepsilon$ limit, we look for a regular expansion of the form
\begin{gather}
\vecu = \vecu^{\sst(0)} + \varepsilon\,\vecu^{\sst(1)} + O(\varepsilon^2),\label{expand_eps_1}\\
p =  p^{\sst(0)} + \varepsilon\,p^{\sst(1)} + O(\varepsilon^2),\label{expand_eps_2}\\
\rho =  \rho^{\sst(0)} + \varepsilon\,\rho^{\sst(1)} + O(\varepsilon^2),\label{expand_eps_3}\\
\vecs = \vecs^{\sst(0)} + \varepsilon\,\vecs^{\sst(1)}
+ O(\varepsilon^2).
\end{gather}
At order one, the system given by equations (\ref{NS11comp}) - (\ref{compressibility}) yields
\begin{gather}
\lambda^2\frac{\partial \vecu^{\sst(0)}}{\partial t} =  \Nab\cdot\vecs^{\sst(0)},\label{expand_order0_1}\\
k_{\sst 0}^2\frac{\partial \rho^{\sst(0)}}{\partial t} + \Nab\cdot\vecu^{\sst(0)} = 0,\label{expand_order0_2}\\
p^{\sst(0)} =  \rho^{\sst(0)}.\label{expand_order0_3}
\end{gather}
Due to the symmetry of the incident field, equation (\ref{expand_lin_incident_flow}), the solution $\vecu^{\sst(0)}$
can be written as the sum of an antisymmetric (uniform) part and a symmetric (linear) part
\begin{equation}
\vecu^{\sst (0)} = \vecu^{\sst(0)}_{\sst A} + \vecu^{\sst(0)}_{\sst S}.\label{sym_asym_vel}
\end{equation}
To order $O(\varepsilon)$, equation (\ref{NS11comp}) then yields
\begin{equation}
\lambda^2\bigg[\rho^{\sst(0)} k_{\sst 0}^2\frac{\partial \vecu^{\sst(0)}}{\partial t} + \frac{\partial \vecu^{\sst(1)}}{\partial t}
+ (\vecu^{\sst(0)}\cdot\Nab)\vecu^{\sst(0)}\bigg] = \Nab\cdot\vecs^{\sst(1)},
\end{equation}
which, when using  equation (\ref{expand_order0_2}) and taking the average in time, leads to
\begin{equation}
\Nab\cdot\langle \vecs^{\sst(1)} \rangle = \lambda^2 \Nab \cdot \langle \vecu^{\sst(0)}\vecu^{\sst(0)} \rangle.
\label{stress_ss}
\end{equation}
Using equation (\ref{sym_asym_vel}), we then obtain
\begin{align}
\Nab\cdot\langle \vecs^{\sst(1)} \rangle = \lambda^2\,[\Nab \cdot \langle \vecu^{\sst(0)}_{\sst A}\vecu^{\sst(0)}_{\sst A}\rangle
+ \Nab \cdot \langle \vecu^{\sst(0)}_{\sst S}&\vecu^{\sst(0)}_{\sst S}\rangle\nonumber\\
+ & \Nab\cdot\langle \vecu^{\sst(0)}_{\sst A}\vecu^{\sst(0)}_{\sst S} +
\vecu^{\sst(0)}_{\sst S}\vecu^{\sst(0)}_{\sst A}
\rangle].
\label{stress_sym_asym}
\end{align}
The first two terms of the right-hand side of equation (\ref{stress_sym_asym}) force a quadrupolar flow
(figure \ref{dip_quad}c) \cite{Riley1966}. In contrast, the last two terms, involving  cross products between $ \vecu^{\sst(0)}_{\sst S}$ 
and $ \vecu^{\sst(0)}_{\sst A}$, give rise to a dipolar flow of axis $\veck_{\sst 0}$ (figure \ref{dip_quad}d) \cite{Danilov2000,Gorkov1962}, 
which, in principle, contributes to the global force experienced by the particle. Considering the respective amplitudes
of $ \vecu^{\sst(0)}_{\sst A}$ and $ \vecu^{\sst(0)}_{\sst S}$, the dipolar 
term is proportional to $k_{\varpi} \sin 2 k_{\sst 0} x_{\sst 0}$, and therefore vanishes for $k_{\sst 0} x_{\sst 0} = \pi/2$, 
which is consistent with the symmetry of the problem of a spherical particle trapped at the nodal pressure plane.
In the general case of a non-spherical particle located at an arbitrary position
in the resonator the dipolar flow should of course be taken into account to derive the transverse drift. 
In the situation considered in this paper, the only remaining net flow is quadrupolar. When the particle is located at
the pressure nodal plane, only the product $\vecu^{\sst(0)}_{\sst A}\vecu^{\sst(0)}_{\sst A}$ 
has a non zero contribution to the quadrupole. The flow $\vecu^{\sst(0)}_{\sst A}$ being uniform,
this is precisely the term  taken into account to assess the transverse velocity of the
particle - {\it i.e.}~the velocity of the particle in the pressure nodal plane, normal to $\veck_{\sst 0}$.

\bibliographystyle{unsrt}
\bibliography{biblio_pr}
\end{document}